\documentclass[12pt]{article}
\newcommand{\be}{\begin{equation}}
\newcommand{\ee}{\end{equation}}
\newcommand{\beq}{\begin{equation}}
\newcommand{\eeq}{\end{equation}}
\newcommand{\bea}{\begin{eqnarray}}
\newcommand{\eea}{\end{eqnarray}}
\newcommand{\nn}{\nonumber}
\newcommand{\lam}{\lambda}
\begin{document}
\baselineskip=15.5pt
\pagestyle{plain}
\setcounter{page}{1}


\def\del{{\partial}}
\def\vev#1{\left\langle #1 \right\rangle}
\def\cn{{\cal N}}
\def\co{{\cal O}}
\newfont{\Bbb}{msbm10 scaled 1200}     
\newcommand{\mathbb}[1]{\mbox{\Bbb #1}}
\def\IC{{\mathbb C}}
\def\IR{{\mathbb R}}
\def\IZ{{\mathbb Z}}
\def\RP{{\bf RP}}
\def\CP{{\bf CP}}
\def\Poincare{{Poincar\'e }}
\def\tr{{\rm tr}}
\def\tp{{\tilde \Phi}}

\def\TL{\hfil$\displaystyle{##}$}
\def\TR{$\displaystyle{{}##}$\hfil}
\def\TC{\hfil$\displaystyle{##}$\hfil}
\def\TT{\hbox{##}}
\def\HLINE{\noalign{\vskip1\jot}\hline\noalign{\vskip1\jot}}
\def\seqalign#1#2{\vcenter{\openup1\jot
  \halign{\strut #1\cr #2 \cr}}}
\def\lbldef#1#2{\expandafter\gdef\csname #1\endcsname {#2}}
\def\eqn#1#2{\lbldef{#1}{(\ref{#1})}%
\begin{equation} #2 \label{#1} \end{equation}}
\def\eqalign#1{\vcenter{\openup1\jot
    \halign{\strut\span\TL & \span\TR\cr #1 \cr
   }}}
\def\eno#1{(\ref{#1})}
\def\href#1#2{#2}
\def\half{{1 \over 2}}

\def\ads{{\it AdS}}
\def\adsp{{\it AdS}$_{p+2}$}
\def\cft{{\it CFT}}

\newcommand{\ber}{\begin{eqnarray}}
\newcommand{\eer}{\end{eqnarray}}

\newcommand{\beqar}{\begin{eqnarray}}
\newcommand{\cN}{{\cal N}}
\newcommand{\cO}{{\cal O}}
\newcommand{\cA}{{\cal A}}
\newcommand{\cT}{{\cal T}}
\newcommand{\cF}{{\cal F}}
\newcommand{\cC}{{\cal C}}
\newcommand{\cR}{{\cal R}}
\newcommand{\cW}{{\cal W}}
\newcommand{\eeqar}{\end{eqnarray}}
\newcommand{\tht}{\thteta}
\newcommand{\lm}{\lambda}\newcommand{\Lm}{\Lambda}
\newcommand{\eps}{\epsilon}


\newcommand{\nonu}{\nonumber}
\newcommand{\oh}{\displaystyle{\frac{1}{2}}}
\newcommand{\dsl}
  {\kern.06em\hbox{\raise.15ex\hbox{$/$}\kern-.56em\hbox{$\partial$}}}
\newcommand{\id}{i\!\!\not\!\partial}
\newcommand{\as}{\not\!\! A}
\newcommand{\ps}{\not\! p}
\newcommand{\ks}{\not\! k}
\newcommand{\D}{{\cal{D}}}
\newcommand{\dv}{d^2x}
\newcommand{\Z}{{\cal Z}}
\newcommand{\N}{{\cal N}}
\newcommand{\Dsl}{\not\!\! D}
\newcommand{\Bsl}{\not\!\! B}
\newcommand{\Psl}{\not\!\! P}
\newcommand{\eeqarr}{\end{eqnarray}}
\newcommand{\ZZ}{{\rm \kern 0.275em Z \kern -0.92em Z}\;}

                                                                                                    
\def\del{{\delta^{\hbox{\sevenrm B}}}} \def\ex{{\hbox{\rm e}}}
\def\azb{A_{\bar z}} \def\az{A_z} \def\bzb{B_{\bar z}} \def\bz{B_z}
\def\czb{C_{\bar z}} \def\cz{C_z} \def\dzb{D_{\bar z}} \def\dz{D_z}
\def\im{{\hbox{\rm Im}}} \def\mod{{\hbox{\rm mod}}} \def\tr{{\hbox{\rm Tr}}}
\def\ch{{\hbox{\rm ch}}} \def\imp{{\hbox{\sevenrm Im}}}
\def\trp{{\hbox{\sevenrm Tr}}} \def\vol{{\hbox{\rm Vol}}}
\def\rl{\Lambda_{\hbox{\sevenrm R}}} \def\wl{\Lambda_{\hbox{\sevenrm W}}}
\def\fc{{\cal F}_{k+\cox}} \def\vev{vacuum expectation value}
\def\nodiv{\mid{\hbox{\hskip-7.8pt/}}}
\def\ie{{\em i.e.}}
\def\ie{\hbox{\it i.e.}}

\def\CC{{\mathchoice
{\rm C\mkern-8mu\vrule height1.45ex depth-.05ex
width.05em\mkern9mu\kern-.05em}
{\rm C\mkern-8mu\vrule height1.45ex depth-.05ex
width.05em\mkern9mu\kern-.05em}
{\rm C\mkern-8mu\vrule height1ex depth-.07ex
width.035em\mkern9mu\kern-.035em}
{\rm C\mkern-8mu\vrule height.65ex depth-.1ex
width.025em\mkern8mu\kern-.025em}}}
                                                                                                    
\def\RR{{\rm I\kern-1.6pt {\rm R}}}
\def\NN{{\rm I\!N}}
\def\ZZ{{\rm Z}\kern-3.8pt {\rm Z} \kern2pt}
\def\IB{\relax{\rm I\kern-.18em B}}
\def\ID{\relax{\rm I\kern-.18em D}}
\def\II{\relax{\rm I\kern-.18em I}}
\def\IP{\relax{\rm I\kern-.18em P}}
\newcommand{\CS}{{\scriptstyle {\rm CS}}}
\newcommand{\CSs}{{\scriptscriptstyle {\rm CS}}}
\newcommand{\rc}{\nonumber\\}
\newcommand{\bear}{\begin{eqnarray}}
\newcommand{\eear}{\end{eqnarray}}
\newcommand{\W}{{\cal W}}
\newcommand{\F}{{\cal F}}
\newcommand{\x}{{\cal O}}
\newcommand{\LL}{{\cal L}}
                                                                                                    
\def\mani{{\cal M}}
\def\calo{{\cal O}}
\def\calb{{\cal B}}
\def\calw{{\cal W}}
\def\calz{{\cal Z}}
\def\cald{{\cal D}}
\def\calc{{\cal C}}
\def\to{\rightarrow}
\def\ele{{\hbox{\sevenrm L}}}
\def\ere{{\hbox{\sevenrm R}}}
\def\zb{{\bar z}}
\def\wb{{\bar w}}
\def\nodiv{\mid{\hbox{\hskip-7.8pt/}}}
\def\menos{\hbox{\hskip-2.9pt}}
\def\dr{\dot R_}
\def\drr{\dot r_}
\def\ds{\dot s_}
\def\da{\dot A_}
\def\dga{\dot \gamma_}
\def\ga{\gamma_}
\def\dal{\dot\alpha_}
\def\al{\alpha_}
\def\cl{{closed}}
\def\cls{{closing}}
\def\vev{vacuum expectation value}
\def\tr{{\rm Tr}}
\def\to{\rightarrow}
\def\too{\longrightarrow}


\def\a{\alpha}
\def\b{\beta}
\def\c{\gamma}
\def\d{\delta}
\def\e{\epsilon}           
\def\f{\phi}               
\def\vf{\varphi}  \def\tvf{\tilde{\varphi}}
\def\vp{\varphi}
\def\g{\gamma}
\def\h{\eta}
\def\i{\iota}
\def\j{\psi}
\def\k{\kappa}                    
\def\l{\lambda}
\def\m{\mu}
\def\n{\nu}
\def\o{\omega}  \def\w{\omega}
\def\q{\theta}  \def\th{\theta}                  
\def\r{\rho}                                     
\def\s{\sigma}                                   
\def\t{\tau}
\def\u{\upsilon}
\def\x{\xi}
\def\z{\zeta}
\def\pt{\tilde{\varphi}}
\def\tt{\tilde{\theta}}
\def\lab{\label}  
\def\6{\partial}
\def\wg{\wedge}
\def\atanh{{\rm arctanh}}
\def\bpsi{\bar{\psi}}
\def\bt{\bar{\theta}}
\def\bvf{\bar{\varphi}}

%
                                                                                                    
\newfont{\namefont}{cmr10}
\newfont{\addfont}{cmti7 scaled 1440}
\newfont{\boldmathfont}{cmbx10}
\newfont{\headfontb}{cmbx10 scaled 1728}
\renewcommand{\theequation}{{\rm\thesection.\arabic{equation}}}
\begin{titlepage}

\begin{center} \Large \bf Dipole Deformations of ${\cal N}=1$ SYM 
and Supergravity backgrounds with $U(1)\times U(1)$ global symmetry 

\end{center}

\vskip 0.3truein
\begin{center}
Umut G{\"u}rsoy ${}^{*}$\footnote{umut@mit.edu} and Carlos 
N\'u\~nez ${}^{*}$\footnote{nunez@lns.mit.edu}
\vspace{0.7in}\\
${}^{*}$ Center for Theoretical Physics, Massachusetts Institute of
Technology \\
Cambridge, MA 02139, USA
\vspace{0.3in}
\end{center}
\vskip 1truein

{\bf ABSTRACT}

We study $SL(3,R)$ deformations of a type IIB background based on $D5$ branes 
that is conjectured to be dual to ${\cal N}=1$ SYM. We argue that this
deformation of the geometry correspond to turning on a dipole
deformation in the field theory on the D5 branes. We give evidence that
this deformation only affects the KK-sector of the dual field theory
and helps decoupling the KK dynamics from the pure gauge dynamics. 
Similar deformations of the geometry that is dual to
${\cal N}=2$ SYM are studied. Finally, we also study  a 
deformation that leaves us with a possible candidate for a dual to $N=0$ 
YM theory.   


\vskip2.6truecm
\vspace{0.3in}
\vspace{0.3in}
\leftline{MIT-CPT 3630}
\leftline{hep-th/0505100 }
\smallskip
\end{titlepage}
\setcounter{footnote}{0}
\tableofcontents

\section{Introduction}

The AdS/CFT conjecture \cite{Maldacena:1997re}
\cite{Gubser:1998bc} \cite{Witten:1998qj} is one of the most 
powerful analytic tools for studying strong coupling effects in gauge 
theories. There are many examples that go beyond the initially 
conjectured duality. First steps in generalizing the original duality to non-conformal examples 
were taken in \cite{Itzhaki:1998dd}. 
Later, very interesting developments led to the construction of the gauge-string 
duality in phenomenologically more relevant theories \ie\, minimally or non-supersymmetric gauge theories 
\cite{Girardello:1999bd}. New geometries that realize various different aspects of gauge theories 
allowed us to deepen our knowledge on the duality. 

Conceptually, the most clear set up for less symmetric 
theories is obtained by breaking the conformality and partial supersymmetry 
by deforming ${\cal N}=4$ SYM with relevant operators or VEV's.  
The models put forward by Polchinski and Strassler 
\cite{Polchinski:2000uf} and Klebanov and 
Strassler \cite{Klebanov:2000hb} belong to this class. Many authors have 
contributed to the understanding of this class of set-ups. 
The review \cite{Skenderis:2002wp}
provides a nice summary of the  procedure to compute 
correlation functions and various observables. 

On the other hand, a different set of models, that are less 
conventional regarding the UV completion of the field theory have been 
developed. The idea here is to start from a set of Dp-branes 
(usually with $p>3$), that 
wrap a q-dimensional compact manifold in a 
way such that two conditions are 
satisfied: One imposes that the low energy description of the 
system is ($p-q$) dimensional, that is, the size of 
the q-manifold is small and is 
not observable at low energies. Secondly, it 
gives technical control over the theory 
to require that a minimal amount of SUSY 
is preserved. For example, the resolution of the 
the Einstein eqs. is eased.  
It turns out that, in this second class of phenomenologically 
interesting dualities, the UV completion of the field theory is 
a higher dimensional field theory.

There are several 
models that belong to this latter class, and in this paper we are 
interested in those that are dual to $N=1$ and $N=2$ SYM. The model dual to $N=1$ SYM 
\cite{Maldacena:2000yy} builds on a 
geometry that was originally found in 4-d gauged supergravity in 
\cite{Chamseddine:1997nm}. The model that is dual to $N=2$ SYM was later found in
\cite{Gauntlett:2001ps}. 
It must be noted that all of the models in this category are afflicted  
by the following problem: they are not dual to ``pure'' field theory of 
interest, but instead, the field theory degrees of 
freedom are entangled with the KK modes on the 
$q$-manifold in a way that depends on the energy scale of the field theory.  
The KK modes enter the theory at an energy scale which is inversely 
proportional to the size of the q-manifold and the main problem 
is that this size is comparable to the 
scale that one wants to study non-perturbative phenomena such as 
confinement, spontaneous breaking of chiral symmetry, etc. 
Nevertheless, this limitation can be seen as an artifact of the supergravity 
approximation and will hopefully be avoided once the formulations 
of the string sigma model on these RR backgrounds becomes available.
Many articles have studied different 
aspects of these models.  Instead of revisiting the main results here, we 
refer the interested reader to the review articles \cite{Bertolini:2003iv}. 

Very recently, an interesting 
development took place by the paper of Lunin and Maldacena 
\cite{Lunin:2005jy}. The authors considered a general background of IIB SG 
\footnote{See \cite{Lunin:2005jy} for discussion also in cases of IIA and 11D SG.}  
that possess two shift isometries, hence includes a torus as a part of the geometry. 
This $U(1)\times U(1)$ isometry allows one to generate new SG solutions by performing 
an $SL(2,R)$ transformation on the $\tau$ parameter of the torus:
\beq
\tau=B_{12}+i\sqrt{\det[g]},
\eeq
where the real and imaginary parts are the component of the NS 
two-form along the torus and the volume of the torus respectively. 
This transformation combined with the usual $SL(2,R)$ symmetry 
of the IIB theory (that acts on the axion and dilaton) 
closes onto a larger group  $SL(3,R)$. A two-parameter subgroup of this 
general symmetry is singled out by the requirement of regularity 
in the transformed solutions. This specific subgroup is referred to as the 
$\beta$-transformation.  A natural question 
from the point of view of the gauge-gravity duality concerns the dual of 
the $\b$ transformations on the field theory side. In other words, 
if we consider a geometry that is associated to a known field theory, what 
deformation in the field theory does the transformed solution produce?    
The answer to this question that was proposed by Lunin and Maldacena 
\cite{Lunin:2005jy} is quite interesting: 
Associated with the $U(1)\times U(1)$ 
isometry of the geometry, there are two separate shift 
transformations that 
act on the component fields of the field theory. 
If we denote the charges of two canonical fields 
$\phi_1$ and $\phi_2$ in the field theory as $q_i^1$ and $q^2_i$ 
under these transformations, then, the effect 
of the $\b$-transformation in the 
dual field theory can be viewed as modifying 
the product of fields in the Lagrangian according to the following rule: 
\beq\lab{qs}
\phi^1(x)\phi^2(x)\to \phi^1(x)*\phi^2(x)= e^{i \pi \beta \det(q_i^j)}\phi^1(x)\phi^2(x),
\eeq 
Therefore this deformation is very much in the spirit of non-commutative deformations 
of field theories \cite{Seiberg:1999vs}. 

The reason behind this result 
lies in the consideration of associated D-brane picture. 
One considers the geometry produced by a number of D-branes. Then the general idea in        
\cite{Lunin:2005jy} is that, depending on the different locations of
the torus in the full geometry, one introduces 
various different type of $\b$-deformations on the gauge theory that lives
on the D-branes. For example, the choice of the torus in the directions transverse 
to the D-branes yields a deformation where the two symmetry transformations in (\ref{qs}) are two global 
$U(1)$ symmetries of the field theory. Lunin and Maldacena gave a specific 
Leigh-Strassler deformation \cite{Leigh:1995ep} of ${\cal N}=4$ SYM 
as an example of this case.    
On the other hand, when the torus is along the D-brane coordinates then 
the associated deformation of the field theory is precisely the standard non-commutative deformation 
of the field theory 
along the torus directions. In this case, the two charges in (\ref{qs}) are the momenta $q^{i}_{x,y} =p^{i}_{x,y}$ of 
$\phi^i$ along the torus. 
Finally, another interesting case that we have more to say in this paper is the case where one 
of the torus directions is along the branes and the other along one of the transverse directions. 
In this case the $\beta$-transformation of the original geometry corresponds to the 
so-called ``dipole deformation'' of the field theory  
\cite{Bergman:2000cw}.  
\subsection{General idea of this paper}
In this work, we consider the ${\cal N}=1$ and ${\cal N}=2$ geometries of
\cite{Maldacena:2000yy} and 
\cite{Gauntlett:2001ps} and study the effects of 
various $\b$-deformations. 
From the general arguments of \cite{Lunin:2005jy}
that we repeated above, we expect that, if one considers a 
toroidal isometry that 
is transverse to the field theory 
directions, and one makes the $\beta$-deformation 
along these directions, this will modify {\em only} the fields 
that are charged under these transverse directions. In other words, in the field theory 
dual to this particular $\b$-transformed theory,  
the dynamics of the KK-modes will 
be modified, whereas the gauge theory dynamics---that we are ultimately 
interested in---will not be affected. Then, one may ask, whether or 
not 
the change that one produces in the KK-sector of the field theory cures the problem of 
entanglement of these unwanted modes with the ``pure'' gauge theory dynamics. 
In this paper we present evidence that the answer is in the affirmative. 
We present our discussion mainly for the case of ${\cal N}=1$ theory, but 
the same considerations apply in the case of ${\cal N}=2$.   

Specifically,
we consider the geometry that is presented in \cite{Maldacena:2000yy}
and apply a {\em{real}} 
$\b$ transformation that also keeps the 
supersymmetry of the theory intact. Then we repeat the computations
 of the VEV of Wilson loops, $\theta_{YM}$ and $\beta_{YM}$-function 
of the theory from the deformed geometry. 
We show that these results are independent of the deformation parameter. 
The case of imaginary $\b$ deformation is also interesting in that it 
changes these results, however we show that 
in that case the dual geometry is singular, 
hence the gravity computations of the field 
theory quantites cannot be trusted.  
In order to investigate whether the 
real $\beta$ transformation affects the KK sector of the theory, 
we compute the masses of a particular 
kind of KK modes in the deformed geometry. This computation is done
both as a dipole field theory computation and as a supergravity
computation. As a field theory calculation it is easy to show that the
particular dipole deformations that we consider when reduced to 4D yield
shifts in the KK masses. In the supergravity side, 
we compute the volume of 
$S^3$ in the deformed solution and show that indeed the volume 
becomes smaller as one turns on $\b$. 
Therefore these  KK-modes indeed 
begin to decouple from the pure gauge theory dynamics 
when one turns on the deformation parameter. 
We also consider the pp-wave 
limits of our deformed geometries with the same goal in mind. 
Namely, to analyze the effects of deformation on the 
dynamics of the KK-sector. The analysis results in a $\beta$ 
correction 
to the original pp-wave that was obtained from 
the original geometry of \cite{Maldacena:2000yy}. 
Therefore the pp-wave analysis also confirms the claim that the 
dynamics 
of the KK-modes are modified in a non-trivial way. 
Similar observations hold for the case of $\b$-deformed 
${\cal N}=2$ geometry. 

We also consider another interesting geometry 
that is obtained from the {\em singular} ``UV'' solution that 
was found in \cite{Maldacena:2000yy}. 
The singular solutions dual to minimally supersymmetric gauge theories 
are interesting in their own right. 
Indeed, historically first the singular dual geometries have been 
found \cite{Klebanov:2000nc}\cite{Maldacena:2000yy}
and then the resolution of the singularities have been discovered 
\cite{Klebanov:2000hb}\cite{Maldacena:2000yy}. 
The singular solutions generally believed to encode 
information on the UV behavior of ${\cal N}=1$ SYM. 
The singular solution in \cite{Maldacena:2000yy} 
preserves an additional $U(1)$ symmetry that is associated 
with the chiral R-symmetry of the ${\cal N}=1$ SYM in the UV. 
It is sometimes denoted as the $\psi$ isometry of the solution. 
We consider choosing the torus such that 
one leg is along the direction of $\psi$. 
Then we perform the $\b$ transformation to generate 
new solutions. As this $SL(3,R)$ transformation does not commute 
with  the R-symmetry the resulting theory is 
non-supersymmetric. Therefore by this method, 
we generate a geometry that would-be a candidate for a dual of 
non-supersymmetric pure YM, once the singularity at $r=0$ is 
resolved. 
We also observe that in the case of $\b$-transforming along the 
R-symmetry direction {\em} both the real and imaginary parts of the 
transformation is allowed: One does not generate 
any irregularities that were present 
in the previous case of supersymmetry preserving transformation. 
This may rise some hope that in 
a particular regime of the relatively 
larger parameter space of the theory 
(now consists of both real and the imaginary parts of $\beta$) 
one may be able to find a resolution to the singularity at the origin.
We leave this question for future work.  
          
In the next section, we move on to 
a presentation of the $\b$ deformations of \cite{Lunin:2005jy}. 
Rather than repeating the 
discussion in \cite{Lunin:2005jy}, we introduce the basic idea and the 
technical aspects of the method in a simple ``warm-up'' example: the 
flat $D5$ geometry. 
We stress the discussion of the 
irregularities associated with the imaginary 
$\b$ transformations. 

Section three reviews the dual of 
${\cal  N}=1$. We review the geometry and present a detailed 
discussion of the mixing between the KK 
and pure gauge dynamics in the original theory. 
Specifically, we review the 
``twisting'' procedure and we make a comparison of confinement and 
KK scales. In this section we also present our field theory argument
for the change in the masses of the KK modes. We show that the dipole
deformation of the 6D field theory when reduced to 4D indeed realizes
the idea of improving the KK entanglement problem.

In section 4, we present the $\b$-transformation 
of the non-singular solution
where the $\b$ transformation is chosen 
in such a way to preserve supersymmetry. 
In this section, we also obtain 
and discuss the aforementioned 
non-supersymmetric solution that 
explicitly breaks the $\psi$-isometry of singular
${\cal N}=1$ solution. 

The main results of our work are presented in section 5, 
that is devoted to the discussion of the $\b$ deformation of the 
non-singular ${\cal N}=1$
geometry. 
There we introduce the transformed solution and discuss 
its properties. Finally we compute the 
field theory observables of interest. 
In particular, we show that the expectation value of the Wilson 
loop, $\theta_{YM}$ and the beta function $\beta_{YM}$ 
are independent of the deformation parameter 
and we present the change in the 
mass of the KK-modes on $S^3$; we also study domain walls and their 
tensions. 

Section 6 discusses pp-wave limits of
some of our solutions. 
We first make some general observations 
about how to perform the pp-wave limits in general 
$\b$-deformed geometries. In particular we argue that 
$\b$ should be scaled to zero along 
with $R\to\infty$. This fact was already observed in 
\cite{Lunin:2005jy}. 
\footnote{Note a misprint in \cite{Lunin:2005jy}: 
$\b$ goes to zero instead of $\infty$.}     
Then we apply this general method of 
taking the pp-waves to three geometries in the following order:
The flat $\beta$-transformed $D5$ geometry, 
the transformed geometry obtained from 
the singular geometry of $D5$ wrapped on $S2$ 
and finally the non-singular $\b$ deformed ${\cal N}=1$
geometry. 

Section 7 includes our results 
for the deformations of the ${\cal N}=2$ geometry. 
We give a summary and discuss various 
open directions in our work in the final section 8. 
Various appendices contain the details of our computations.  

Here is a brief summary of the very recent literature on the subject. 
After the paper \cite{Lunin:2005jy} presented the idea described 
above, together with some checks,
Ref. \cite{Frolov:2005ty} studied the 
proposal from the view point of comparing semi-classical strings 
moving in these backgrounds with anomalous dimensions of gauge 
theory operators. Also aspects of integrability of the spin chain
system that is associated with the Leigh-Strassler deformation are 
studied.
Ref. \cite{Frolov:2005dj}, presented a nice way of understanding 
(the bosonic NS part of) the $SL(3,R)$ transformations in terms of 
T dualities; also made a connection with Lax pairs and found new 
non-supersymmetric deformations.

A final note on the notation: 
As we discuss in the next section, we will mainly be concerned with
the {\em real} $\beta$ transformations in this paper ---unless specified
otherwise--- because of the concerns about irregularity of the 
imaginary $\beta$ transformations. These specific real transformations 
were
denoted as $\gamma$-transformations in \cite{Lunin:2005jy}. Therefore
the term $\gamma$-transformation will be used instead of $\b$ in what
follows.   

\section{Warm-up: Transformations of the Flat D5 Brane Solution}

In this section, we outline the solution generating technique in a
simple warm-up example. This allows us to introduce the basic idea and
the necessary notation that will be used in the following sections. More
importantly, the examples we discuss here shall serve as an illustration
of when and how the $SL(3,R)$ transformations lead to irregular
solutions. The criteria for regularity of the transformation was
outlined in \cite{Lunin:2005jy} and a specific transformation of the 
NS5 (or D5) brane was mentioned as an example of irregular behavior.
Here, we discuss this point in detail. 

\subsection{The flat D5 brane}

Let us consider $N$ D5 branes in flat 10D space-time. The solution in
the string frame is,  
\be\lab{D5sol}
  ds^2 = e^{\f}\left[dx_{1,5}^2 +\a'g_s 
N(dr^2+\frac{1}{4}\sum_{i=1}^3 w_i^2)\right],\; F_{(3)}  = \frac{\a'N}{4}
  \w_1\wg w_2\wg w_3,\; e^{\f}=\frac{\a'g_s (2\pi)^{\frac32}}{\sqrt{N}}e^r.
\ee
We define the $su(2)$ left-invariant one forms as,
\bea\lab{su2}
w_1&=& \cos\psi d\tilde\theta\,+\,\sin\psi\sin\tilde\theta
d\tilde\varphi\,\,,\rc\rc
w_2&=&-\sin\psi d\tilde\theta\,+\,\cos\psi\sin\tilde\theta
d\tilde\varphi\,\,,\rc\rc
w_3&=&d\psi\,+\,\cos\tilde\theta d\tilde\varphi\,\,.
\eea
Ranges of the three angles are $0\le\tilde\varphi< 2\pi$, $0\le\tilde\theta\le\pi$ and
$0\le\psi< 4\pi$. 

One easily sees that this solution includes a torus that is 
parametrized by $\psi$ and $\pt$ as part of its isometries. In order
to perform the $SL(3,R)$ transformation that was specified in
\cite{Lunin:2005jy}, one writes (\ref{D5sol}) in the form that
separates the torus part from the 8D part that is transverse to the
torus. The 8D part is left invariant (up to an overall factor) under
the transformation. We shall use the notation introduced in
\cite{Lunin:2005jy} in order to avoid confusion (see our Appendix 
A for a summary). The D5 metric,
(\ref{D5sol}) can be written as, 
\be\lab{genmetric}
ds^2 = \frac{F}{\sqrt{\Delta}}\left(D\vf_1-C D\vf_2\right)^2+
F\sqrt{\Delta}(D\vf_2)^2 + g_{\m\n}dx^{\m}dx^{\n}.
\ee
In this particular case, the torus is given by,
\be\lab{ofsD5}
D\vf_1 = d\psi + {\cal A}^{(1)},\; D\vf_2 = d\pt + {\cal A}^{(2)}.
\ee
The connection one-forms, ${\cal A}^{(1)}$ and ${\cal A}^{(2)}$ that are required
to put a generic metric in
the above form, vanish in this simple case:
\be\lab{As}
{\cal A}^{(1)}={\cal A}^{(2)}=0.
\ee
The $8D$ part of the metric in (\ref{D5sol}) is,
\be\lab{8DD5} 
g_{\m\n}dx^{\m}dx^{\n} = dx_{1,5}^2 +N(dr^2+\frac{1}{4}d\tt^2).
\ee
We introduced the following metric functions,
ˆ\be\lab{metfuncD5}
  F = \frac{\a' g_s N}{4}e^{\f}\sin\tt,\; \Delta=\sin^2\tt,\; C = -\cos\tt.
\ee 

Similarly, one separates the torus and the transverse part in the
RR two form in a generic way as follows \cite{Lunin:2005jy} and 
Appendix A:
\be
C_2 = C_{12} D\vf_1 \wedge D\vf_2 
+ C^{(1)} \wg D\vf_1 + C^{(2)} \wg D\vf_2 -\frac{1}{2}({\cal A}^{(a)}\wg
C^{(a)} - \tilde{c}).
\label{c2} 
\ee
$C^{(1)}$ and $C^{(2)}
$ are one-forms and $\tilde{c}$ is a two-form on
the 8D transverse part. In this particular example, one can take the
RR form as follows:
\be\lab{NS1}
'
C_2= \frac{\a' N}{4}\psi w_1\wg w_2.
\ee
Then, various objects that appear in 
the general formula (\ref{c2}) become, 
\be\lab{c2D5}
C^{(2)} = \frac{\a' N}{4}\psi\sin\tt d\tt,\; C_{12}=C^{(1)}=\tilde{c} = 0.
\ee

\subsection{The General Transformation}

Now, we consider the transformation of a general solution of IIB SG under the 
following two-parameter subgroup of $SL(3,R)$ \footnote{This specific
  subgroup - actually a larger one which also includes the ordinary
  $SL(2,R)$ of IIB in part - was specified in \cite{Lunin:2005jy} as
  a necessary condition for the regularity of the transformed
  geometry. Although necessary, it is not sufficient for the
  regularity, as we discuss further below.}: 
\begin{displaymath}\lab{SL}
\Lambda = \left(\begin{array}{ccc}
1 & \g & 0\\
0 & 1  & 0\\
0 & \s & 1
\end{array}\right)
\end{displaymath}
In the next subsection, we work out the details of the transformation
in the simple example of (\ref{D5sol}). 

Consider a general solution to IIB Supergravity.  
Various components of the RR two-form, NSNS two-form, the four-form and the $\cA$-vectors
that are defined in (\ref{genmetric}) are
grouped into the following combinations  that
transform as vectors under $SL(3,R)$  
(\cite{Lunin:2005jy} and 
Appendix A):
\bea
& & V^{(i)}_\mu =(-\epsilon^{ij} B^{(j)}_\mu,A^{(i)}_\mu,\epsilon^{ij} 
C^{(j)}_\mu),\;\;\nonumber\\ 
& & W_{\mu\nu} =(\tilde{c}_{\mu\nu}, \tilde{d}_{\mu\nu},\tilde{b}_{\mu\nu}).
\label{transf11}
\eea
Here $B^{(1)}$, $B^{(2)}$, 
$\tilde{b}$ and $\tilde{d}$ are components of the NS
form and the four-form that follows from the separation of the torus
and the transverse part in complete analogy with 
(\ref{c2})\cite{Lunin:2005jy}.  
Their explicit transformation under (\ref{SL}) is given as, 
\bea
& & (V^{(i)})'= (- \epsilon^{ij} B^{(j)}, 
A^{(i)} + \gamma \epsilon^{ij} B^{(j)} - \sigma 
\epsilon^{ij} C^{(j)}, \epsilon^{ij} C^{(j)})
\nonumber\\ 
& & W'= (\tilde{c} +\gamma \tilde{d}, \tilde{d}, \tilde{b} +\sigma \tilde{d})
\label{vectors}
\eea

Transformation of various scalar fields that we defined above is
obtained as follows (see our Appendix 
A for a summary of these results). 
One constructs a $3\times 3$ matrix, $g^T$, from the 
scalar $F$, the dilaton, and the  torus components of
the RR and NS forms, $C_{12}$ and $B_{12}$.   
The components of the initial $g^T$ are \footnote{Here, we consider
  the special case of $B_{12}=0$. The most general case is further
  discussed in the Appendix A.},
\beq
g^T_{1,1}=  (e^\phi F)^{-1/3}, \;\; g^T_{2,2} =e^{-\phi/3} F^{2/3}, 
\;\;g^T_{3,2}= - C_{12} (F)^{-1/3} e^{2/3\phi}, \;\; g^T_{3,3}=e^{2/3\phi} (F)^{-1/3}
\lab{comps}
\eeq
and all of the other components are zero. 

Now, a few words about how to obtain the transformed matrix: The
object that transforms as a matrix under $SL(3,R)$ is the following
one:
\be 
M = g g^T, \;\;\; M\to \Lambda M \Lambda^T.  
\ee
Thus, one can read off the transformation of $g^T$ as, 
\be
g^T \to \eta^T g^T\Lambda^T,
\ee
where $\eta$ is an $SO(3)$ matrix that can be parametrized by three
Euler angles. These angles are determined by demanding that the
transformed $g^T$ has the specific structure given in
\cite{Lunin:2005jy}, namely its $(1,3), (2,1)$ and $(2,3)$ components
vanish. 
We obtain the new values for $F, e^{\phi}, C_{12}, \chi, 
B_{12}$ as,
\beq
\lab{gentrans}
F'= F G\sqrt{H},\;\; e^{\phi'}=e^{\phi}H \sqrt{G},\;\; B_{12}'=\gamma
F^2 G,\;\; \chi'=\gamma \frac{J}{H},\;\; C_{12}'=-JG
\eeq
where we defined, 
\bea\label{GHJ}
H & = & (1-C_{12}\sigma)^2+F^2\sigma^2e^{-2\phi},\nonumber\\
 G & = & ((1 -
C_{12}\sigma)^2+F^2\sigma^2e^{-2\phi}+\gamma^2F^2)^{-1},\\
 J & = & \sigma
F^2e^{-2\phi}-C_{12} (1-C_{12}\sigma).\nonumber 
\eea

Using these transformation properties 
one obtains the new metric as follows:
\be\lab{newmetric}
ds^2 = \frac{F'}{\sqrt{D}}\left(D\vf_1'-C D\vf_2'\right)^2+
F'\sqrt{D}(D\vf_2')^2 + U_{st} g_{\m\n}dx^{\m}dx^{\n}.
\label{transf18}\ee
Here $D\vf_i'$ include the transformed $\cA$ forms, \ie $ D\vf_i' =
d\vf_i + {\cA'}^{(i)}$. The volume ratio that appears in front of the
8D transverse part is, 
\be\lab{ust}
U_{st}  = \frac{e^{\frac23(\f'-\f)}}{(\frac{F'}{F})^{\frac13}} = H^{\half}.
\ee
where we used (\ref{gentrans}) in the last line. The expression for
$H$ in (\ref{GHJ}) tells us the important fact that {\em the 8D volume
  ratio is different than 1 only for non-zero $\s$
  transformations}.
In particular an arbitrary $\g$ transformation (with $\s=0$), leaves 
the
8D volume invariant:
\be\lab{ust1}
U_{st}(\s=0) = 1.
\ee

To extract useful information about the field theory that is dual to
the transformed geometry, one often needs the analogous expression in
the Einstein frame. Making a Weyl transformation to the Einstein
frame, one obtains the following volume ratio of the 8D part in the
Einstein frame: 
\be\lab{ue}
U_E = e^{-\half(\f'-\f)}U_{st} = G^{-\frac14}.
\ee
This ratio is generally different than one for any transformation. 
    
\subsection{Regularity of Transformed D5}

Now, let us apply this procedure to the particular solution given in
(\ref{metfuncD5}) and (\ref{c2D5}). From (\ref{vectors}) we read off the
  new values of the $\cA$ forms, and the vector components of the RR
  form as, 
\bea\lab{newvecs}
{{\cA}^{(1)}}' & = & - \s C^{(2)} = -\s \frac{\a' N}{4} \psi \sin\tt d\tt,\\ 
{C^{(2)}}' & = & C^{(2)} = -\frac{\a'N}{4} \psi \sin\tt d\tt,\; 
{{\cA}^{(2)}}' = {C^{(1)}}' =0\nonumber
\eea

Various scalar fields in the new solution are obtained from
(\ref{gentrans}). The new metric is, using (\ref{newvecs}) in (\ref{newmetric}), 
 \be\lab{newD5}
ds^2 = \frac{F'}{\sqrt{\Delta}}\left(d\psi - \s \frac{\a'N}{4} 
\psi \sin\tt d\tt + \cos\tt d\pt^2\right)^2+
F'\sqrt{\Delta}d\pt^2 + U_{st} g_{\m\n}dx^{\m}dx^{\n}.
\ee
The particular appearance of $\psi$ above makes the transformed metric
irregular. This is because, $\psi$ originally was defined as a periodic
variable with period $4\pi$. However the transformed metric is no
longer periodic in $\psi$. One can easily track the origin of this
irregularity: It is coming from the contribution of the $\s$
transformation to the $\cA$ one-forms in (\ref{newvecs}). In
this case where the torus is chosen in the directions transverse to the D5
brane, the RR two form has the particular form in (\ref{c2D5}) with a
bare dependence on $\psi$ and this bare dependence directly carries on
to the metric under the $\s$ transformation. Notice that this irregular behavior does
not happen for a one parameter $\g$-transformation where one sets
$\s=0$. In that case one obtains a new regular solution to IIB SG! (or at 
least one in which we do not generate {\it new} singularities). 

One may wonder  if this irregularity is due to our particular
choice of (\ref{NS1}): $C_2$ is defined only up to a gauge
transformation and one can try to use a gauge-equivalent expression
where $\psi$ does not appear in this form. For example a gauge
equivalent choice of $C_2$ is given by,
\be\lab{c22}
C_2 = \frac{\a'N}{4} d\psi\wg w_3 = \frac{\a'N}{4} \cos\tt d\psi\wg d\pt.
\ee
Comparison with the general expression, (\ref{c2}) shows that,
\be\lab{c222}
C_{12} = \frac{\a'N}{4} \cos\tt,
\ee     
and the rest of the components in (\ref{c2}) vanish. In particular,
the vector $C^{(1)}$ vanish and one does not generate any irregular
behavior in the metric, as in (\ref{newD5}). 

However, as described in
\cite{Lunin:2005jy}, for the regularity of the full solution, 
one should also make sure that the RR two-form in the original
solution go to the same integer at various potentially
dangerous points where the volume of the torus shrinks to
zero size.\footnote{In \cite{Lunin:2005jy}, this argument was made for
  $B_2$ under the $\g$ transformation. Same argument applies to $C_2$
  in the case of $\s$ transformation.} One sees from
(\ref{D5sol}) that these singular points where the volume of the
$\psi$-$\pt$ torus shrinks to zero are given by
$\tt=0$ and $\pi$. Equation, (\ref{c222}) tells us that at these
points, $C_2$ goes over to
$-N$ and $+N$, respectively. As $\tt$ is a
periodic variable with period $\pi$, a discrete jump of $2N$ in the
flux as one completes the period is unacceptable and generates an infinite field 
strength. Thus we conclude
that, in case where the torus is chosen in the transverse directions to the D5
brane, $\s$ transformation is sick, independently of the gauge choice for
$C_2$. This does not happen for the $\g$ transformation. With a
completely analogous argument, one shows that \footnote{One can use the
  formulae given in the previous subsection to work this case
  out. Our formula (\ref{gentrans}) is applicable only to the
  case $B_{12}=0$ (but we wrote general formulas in Appendix A) and one can choose 
a gauge in the NS5 solution such
  that this happens.}  the reverse phenomena
happens in case of the NS5 brane solution. In that case there is a
non-trivial $B_2$ form and this time the $\g$ transformation exhibits
the same irregular behavior, whereas the $\s$ transformation is free
of irregularity.  

Finally, let us recall that the Ricci scalar of the original 
flat D5 brane (\ref{D5sol}) geometry is bounded for large values 
of the radial coordinate, 
where the dilaton becomes large. The divergence in the dilaton 
indicates the need to passing to an S-dual description that is given
 in terms of NS5 branes. At small values of the radial coordinate, 
the Ricci scalar diverges instead, thus indicating that we are in 
the regime where the 6D SYM theory is the weakly coupled description 
of the system.
In the case of the transformed metric (\ref{newmetric}) things are 
more interesting. One can check that, for large values of 
the radial coordinate, the transformed dilaton in (\ref{gentrans}) 
does not diverge (except for the values $\theta =0,\; \theta=\pi$) 
and the Ricci scalar has an expression that depends on the
transformation parameter, 
\beq
R_{eff} \approx e^{-r}\Big(\frac{-384 + 26\eta^2\gamma^2 e^{2r} 
-\eta^2\gamma^4 
e^{4r} +\eta^2\gamma^2 \cos(2\theta)e^{2r}(54 + e^{2r} 
\gamma^2)}{\eta^2(16 
+\gamma^2 
e^{2r}\sin^2\theta)} \Big), \eta=\sqrt{g_{YM}^2 N (2\pi)^{-3}}, 
\label{newcurvature}
\eeq
where $\eta=\sqrt{g_{YM}^2 N (2\pi)^{-3}}$.  
So, if we fix $\theta$ as $n\pi$, for large values of the 
radial coordinate, we should pass to a 
dual description. In this case, doing a T-duality seems appropriate.
On the other hand, for values of the angle $\theta$ different from 
$n\pi$ we see that the Ricci scalar (which is non-zero) and the 
dilaton are bounded for 
large values of the radial coordinates, in contrast with the usual 
$D5$ case. For this case, 
it does not seem necessary to pass to a NS5 description.

For values of the radial coordinate near negative infinity, the 
Ricci scalar 
diverges and the good description is in terms of a 6D gauge 
theory. The main point of the discussion above is the fact that the 
transformed system has in 
principle a more complicated `phase space' than the original flat 
D5-branes \cite{Itzhaki:1998dd}.

\subsection{The Gauge Theory}

Let us briefly comment on the gauge theory dual of the transformed
$D5$ background. 
First, let us recall that the gauge theory on flat D5 branes 
is a 6D maximally supersymmetric Yang-Mills theory. The bosonic part of the 
Lagrangian can be obtained by reducing ten-dimensional SYM on a  
four-torus. Schematically, it is given as
\beq\lab{D5lagr}
S=Tr\int d^{6}x\Big( F_{\mu\nu}^2 + (D_\mu\Phi)^2 + 
V([\Phi,\Phi])\Big)
\eeq

According to the prescription of \cite{Lunin:2005jy} the new
background in (\ref{transf18}) is dual to (\ref{D5lagr}) but with the
potential replaced by a $\g$ deformed one obtained by replacing the
products of fields in $V[\dots]$ by the deformed product of
(\ref{qs}). We would like to remark that as the torus of
transformation that is formed by $\psi$ and $\tilde{\vf}$ is
transverse to the D5 branes this will only introduce some phases in
the scalar potential.  
 
One should also note that as the transformation that we perform
mixes angles that correspond to the $SU(2)_L\times SU(2)_R$ R-symmetry 
of the 6D SYM theory, this transformation will break supersymmetry.
We shall not dwell on this issue for this warm-up exercise but it is
discussed in more detail for the model of our interest in the next
section. 

\subsection{Dual of the Dipole Theory}

We would like to end this warm-up section by a discussion of the
dipole theories that shall interest us in the following
sections. Our main interest in these exotic quantum theories lies in the 
fact that the transformations studied in this section when 
applied to a supergravity dual to ${\cal N}=1$ SYM leads to a dipole 
theory for the KK section of the field theory. This is discussed in
section 3.2 in detail. For the literature on the dipole theories, 
see \cite{Bergman:2000cw}. Here we only summarize some features that
will interest us. 

A dipole theory is a field theory where the locality is lost due to the 
fact that some of the fields come equipped with a ``dipolar moment'' 
$L_\mu$ such that the product of  fields $(\Phi_1, \Phi_2)$ with 
moments $\vec{L}_1, \vec{L}_2$ is given by
\beq
\Phi_1(x) \Phi_2(x)= \Phi_1(x-L_2/2) \Phi_2(x+L_1/2)
\eeq
The non-locality of the interaction is manifest. Also, the 
presence of dipole moments break Lorentz invariance. We are 
interested in the particular case where all of 
the dipole moments point in the same direction.

If we have a field theory with many fields $\Phi_i$, each one
of them with a global conserved charge $q_i$, we can pick an 
arbitrary vector $L_\mu$ and the dipole moment of each field is taken
as $q_i L_\mu$. More generally, if for each field $\Phi_i$ 
in a collection of $n$ fields, there exist 
a set of conserved charges $q^i_1, ...q^i_k$, then one can pick an
 $n\times k$ matrix (that we call $S$) and assign
each field a 
dipole moment of the form $\vec{L}=S \vec{q}$. In this general case 
the arbitrary choice of $S$, breaks the Lorentz invariance.

The way to construct the dipole Lagrangian is clearly explained in 
\cite{Bergman:2000cw}. Basically, the idea follows the approach of 
\cite{Seiberg:1999vs}, only that in this case, we allow for a
non-constant NS field, in addition to the RR forms.
Briefly, the proposal is that, any term in the original 
Lagrangian that is of the following form,
\beq
L_{int}= \Phi_1(x)....\Phi_n(x)
\eeq
is replaced (in momentum space) by
\beq
L'_{int}= e^{\sum_{i<j}^n p_i L_j}\Phi_1(x)....\Phi_n(x)
\eeq
If we follow the proposal of Lunin and Maldacena \cite{Lunin:2005jy} 
that was  reviewed above, this is equivalent to performing an $SL(3,R)$ 
transformation on a torus that has one direction on the brane and one 
external to the brane. 

To improve and clarify this discussion, 
let us now construct the background dual to a dipole 6D
field theory. Once again, we do not worry about preservation 
of supersymmetry in this section.  
If one performs the $\g$ transformation on the torus with
one direction along the D5 brane and one transverse to it, one produces 
a background dual to a six dimensional dipole theory as described 
above. Obviously the 
Lorentz invariance $SO(1,5)$ is explicitly broken by the choice 
of one coordinate along the brane. The background 
dual to this field theory is given by
\bea
 ds^2 &=& \frac{F'}{\sqrt{\Delta}}(d\psi + \cos\theta d\varphi)^2 + 
F'\sqrt{\Delta}dz^2 + U_{st}\Big(dx_{1,4}^2  +\nonumber\\ 
{}& & \alpha' g_s N (dr^2 + 
\frac{1}{4}(d\theta^2 + 
\sin\theta^2 d\varphi^2))   \Big) 
\label{dipolod5}
\eea
where we labeled by $z, \psi$ the directions of the two torus. 
Following the notation in eq.(\ref{genmetric}), the functions
are defined as 
\beq
F^2= \frac{\alpha' g_s N e^{2\phi}}{4}, \;\;\; \Delta= 
\frac{4}{\alpha' g_s N}.\;\; F'=\frac{F}{1+\gamma^2 F^2}, \;\;{\cal 
A}^{(1)}= \cos\theta d\varphi, {\cal A}^{2}= C^{i}= C=0
\eeq
and $U_st$ is given by (\ref{ust}). 
There is also a dilaton and NS and RR forms that transform 
according to eq.(\ref{gentrans}). The 
original RR two form only has the following component,
\beq
\tilde{c}=2\psi \alpha' N \sin\theta d\theta \wedge d\varphi
\eeq
and it transforms as we indicated in (\ref{vectors}).

Let us briefly comment on the geometry (\ref{dipolod5}). The 
transformed 
dilaton is bounded above for large values of the radial 
coordinate (and for any values of the angles) 
unlike the case of the flat D5 
brane analyzed in the previous subsection. The effective 
curvature is given by, 
\beq
R\approx 2e^{-r}\Big(\frac{-96 + 8\eta^2\gamma^2 e^{2r} +\eta^4 
\gamma^4 e^{4r}}{\eta^2(4 +\gamma^2 e^{2r})^2}\Big), \;\;\; 
\eta=\sqrt{g_{YM}^2 N (2\pi)^{-3}}.
\eeq
Again, we find no problem for the large-r regime, neither for the Ricci 
scalar, nor for the dilaton. It is peculiar that it is possible to 
find a value of the radial coordinate where the Ricci scalar 
vanishes.

\section{${\cal N}=1$ SYM and the KK-Mixing Problem}

\subsection{Review of the Geometry Dual to ${\cal N}=1$ SYM} 

We work with the model 
presented in \cite{Maldacena:2000yy}
(the solution was first found in a 
4d context in \cite{Chamseddine:1997nm}) 
and described and studied in more detail in \cite{Nunez:2003cf}. 
Let us briefly describe the 
main points of this supergravity dual to $N=1$ SYM and its UV completion.

Suppose that we start with N $D5$ branes. 
The field theory that lives on them is 
6D SYM with 16 supercharges. Then, 
suppose that we wrap two directions of the 
D5 branes on a curved two manifold 
that can be chosen as a sphere.
In order to preserve SUSY a twisting procedure 
has to be implemented and 
actually, there are two ways of doing it. 
The one we will be interested in this 
section, deals with a twisting that preserves 
four supercharges. In this case 
the two-cycle mentioned above lives inside a CY3 fold. On the other hand, if 
the twist that preserves eight supercharges is performed, the two cycle lives
inside a CY2-fold 
\cite{Gauntlett:2001ps}. This second case will be analyzed in section
7. We note that this supergravity solution is dual to a 
four dimensional field theory, only for low energies (small values of the 
radial coordinate). 
Indeed, at high energies, the modes of the gauge theory 
begins to fluctuate also on the two-cycle and as the energy is
increased further, the theory first becomes ${\cal N}=1$ SYM in six 
dimensions and then, blowing-up of the dilaton forces one to
S-dualize. Therefore the UV completion of the model is given by 
the little string theory. 

The supergravity solution that interests us in this 
section, preserves four supercharges and has the topology 
$R^{1,3}\times R\times S^2\times S^3$. There is a fibration 
of the two spheres in such a way that that ${\cal N}=1$ supersymmetry
is preserved. By going near $r=0$ it can be seen that the 
topology is $R^{1,6} \times S^3$. The full solution and Killing 
spinors are written in 
detail in \cite{Nunez:2003cf}.
The metric in the Einstein frame reads,
\beq
ds^2_{10}\,=\,\alpha' g_s N e^{{\phi\over 2}}\,\,\Big[\,
\frac{1}{\alpha' g_s N }dx^2_{1,3}\,+\,e^{2h}\,\big(\,d\theta^2+\sin^2\theta 
d\varphi^2\,\big)\,+\,
dr^2\,+\,{1\over 4}\,(w^i-A^i)^2\,\Big]\,\,,
\label{metric}
\eeq
where $\phi$ is the dilaton. The angles
$\theta\in [0,\pi]$ and
$\varphi\in [0,2\pi)$ parametrize a two-sphere. This sphere is fibered in the ten 
dimensional metric by the one-forms
$A^i$ $(i=1,2,3)$.
Their are given in terms of a 
function
$a(r)$ and the angles $(\theta,\varphi)$ as follows:
\beq
A^1\,=\,-a(r) d\theta\,,
\,\,\,\,\,\,\,\,\,
A^2\,=\,a(r) \sin\theta d\varphi\,,
\,\,\,\,\,\,\,\,\,
A^3\,=\,- \cos\theta d\varphi\,.
\label{oneform}
\eeq
The $w^i\,$ one-forms are defined in (\ref{su2}).

The geometry in (\ref{metric}) preserves supersymmetry 
when the functions $a(r)$, $h(r)$ and the dilaton $\phi$ are:
\bea
a(r)&=&{2r\over \sinh 2r}\,\,,\rc\rc
e^{2h}&=&r\coth 2r\,-\,{r^2\over \sinh^2 2r}\,-\,
{1\over 4}\,\,,\rc
e^{-2\phi}&=&e^{-2\phi_0}{2e^h\over \sinh 2r}\,\,,
\label{MNsol}
\eea
where $\phi_0$ is the value of the dilaton at $r=0$. Near the origin $r=0$ the 
function
$e^{2h}$ behaves as $e^{2h}\sim r^2$ and the metric is non-singular. The solution of 
the
type IIB supergravity includes a
RR three-form $F_{(3)}$ that is given by
\beq
\frac{1}{\alpha' N} F_{(3)}\,=\,-{1\over 4}\,\big(\,w^1-A^1\,\big)\wedge
\big(\,w^2-A^2\,\big)\wedge \big(\,w^3-A^3\,\big)\,+\,{1\over 4}\,\,
\sum_a\,F^a\wedge \big(\,w^a-A^a\,\big)\,\,,
\label{RRthreeform}
\eeq
where $F^a$ is the field strength of the su(2) gauge field $A^a$, defined as:
\beq
F^a\,=\,dA^a\,+\,{1\over 2}\epsilon_{abc}\,A^b\wedge A^c\,\,.
\label{fieldstrenght}
\eeq
Different components of $F^a$ read,
                                                                                                    
\beq
F^1\,=\,-a'\,dr\wedge d\theta\,\,,
\,\,\,\,\,\,\,\,\,\,
F^2\,=\,a'\sin\theta dr\wedge d\varphi\,\,,
\,\,\,\,\,\,\,\,\,\,
F^3\,=\,(\,1-a^2\,)\,\sin\theta d\theta\wedge d\varphi\,\,,
\eeq
where the prime denotes derivative with respect to $r$.
Since $dF_{(3)}=0$, one can represent $F_{(3)}$ in terms of a two-form potential
$C_{(2)}$ as $F_{(3)}\,=\,dC_{(2)}$. Actually, it is not difficult to verify that
$C_{(2)}$ can be taken as:
\bea
\frac{C_{(2)}}{\a'N}&=&{1\over 4}\,\Big[\,\psi\,(\,\sin\theta d\theta\wedge d\varphi\,-\,
\sin\tilde\theta d\tilde\theta\wedge d\tilde\varphi\,)
\,-\,\cos\theta\cos\tilde\theta d\varphi\wedge d\tilde\varphi\,-\rc\rc
&&-a\,(\,d\theta\wedge w^1\,-\,\sin\theta d\varphi\wedge w^2\,)\,\Big]\,\,.
\label{RR}
\eea
The equation of motion of $F_{(3)}$ in the Einstein frame is
$d\Big(\,e^{\phi}\,{}^*F_{(3)}\,\Big)=0$, where $*$ denotes Hodge duality. 
Let us  stress that the configuration presented above is non-singular.
Finally, let us mention that the BPS equations also admit a solution in which the function
$a(r)$ vanishes, \ie\ in which the one-form $A^i$ has only one non-vanishing
component, namely $A^{3}$. We will refer to this 
solution as the abelian ${\cal N}=1$
background. Its explicit form can easily be obtained by taking the
$r\rightarrow\infty$ limit of 
the functions given in eq. (\ref{MNsol}). Notice that,
indeed $a(r)\rightarrow 0$ as $r\rightarrow\infty$ in  eq. (\ref{MNsol}).
Neglecting exponentially suppressed terms, one obtains
\beq
e^{2h}\,=\,r\,-\,{1\over 4}\,\,,
\,\,\,\,\,\,\,\,\,\,\,\,\,\,\,\,\,\,(a=0)\,\,,
\eeq
while $\phi$ can be obtained 
from the last equation in (\ref{MNsol}). The metric
of the abelian background is singular at $r=1/4$ (the position 
of the singularity can
be moved to $r=0$ by a redefinition of the radial coordinate). 
This IR singularity of
the abelian background is removed in 
the non-abelian metric by switching on the $A^1,
A^2$ components of the one-form (\ref{oneform}). 

\subsection{Dual Field Theory and the Dipole Deformation of the KK Sector}

Let us first summarize some aspects of the field theory that is 
dual to the  geometry above. 
In \cite{Maldacena:2000yy}
this solution was argued to 
be dual to ${\cal N}=1$ SYM.

The 4D field theory is obtained by reduction of $N$ $D5$ branes on
$S^2$ with a twist that we explain below. Therefore, as the energy
scale of the 4D field theory becomes comparable to the inverse volume
of $S^2$, the KK modes begin to enter the spectrum. 

To analyze the spectrum in more detail, we briefly review the twisting
procedure. In order to have a supersymmetric theory on a curved
manifold like the $S^2$ here, one needs globally defined spinors. A
way to achieve this was introduced in \cite{Witten:1988ze}. In 
our case the
argument goes as follows. As D5 branes wrap the two sphere, the
Lorentz symmetry along the branes decompose as $SO(1,3)\times 
SO(2)$. 
There is also an $SU(2)_L\times SU(2)_R$ symmetry  
that rotates the transverse coordinates. This symmetry 
corresponds to the R-symmetry of the supercharges on the field theory
of D5 branes. One can properly define ${\cal N}=1$ supersymmetry
 on the curved space that is obtained by wrapping the D5 branes on the
 two-cycle, by identifying a $U(1)$ subgroup of either
 $SU(2)_L$ or $SU(2)_R$ R-symmetry with the $SO(2)$ of the
 two-sphere.\footnote{The choice of a diagonal $U(1)$ inside
$SU(2)_L \times SU(2)_R$ leads to an $N=2$ field 
theory instead, see \cite{Gauntlett:2001ps}.}
To fix the notation, let us choose the $U(1)$ in $SU(2)_L$. Having done the
identification with $SO(2)$ of the sphere, we denote this twisted
$U(1)$ as $U(1)_T$. 
 
After this twisting procedure is performed, the fields in the theory
are labeled by the quantum numbers of $SO(1,3)\times U(1)_T\times
SU(2)_R$. The bosonic fields are,
\beq
A^{a}_{\mu}= (4, 0, 1), \;\; \Phi^a=(1, \pm, 1), \;\; 
\xi^{a}=  (1, \pm, 2). \;\;
\label{twistboson}\eeq
Respectively they are the gluon, two 
massive scalars that are coming from the reduction of the original 
6D gauge field on $S^2$, (explicitly from the $A_{\vf}$ and $A_{\q}$ components)   
and finally four other massive scalars 
(that originally represented the positions of the D5 branes in the
transverse $R^4$). As a general rule, all the fields that transform
under $U(1)_T$---the second entry in the above charge 
designation---are massive. For the fermions one has, 
\beq
\lambda^a=(2,0,1),\,(\bar{2},0,1),\;\; 
\Psi^a=(2,++,1),\,(\bar{2},--,1),\;\; 
\psi^a= (2,+,2), \, (\bar{2},-,2).
\label{twistfermion}
\eeq
These fields are the gluino plus some massive fermions 
whose $U(1)_T$ quantum number is 
non-zero. The KK modes in the 4D theory are obtained by the harmonic
decomposition of the massive modes, $\Phi,\xi,\Psi$ and $\psi$ that
are shown above. Their mass is of the order of 
$M_{KK}^2= (Vol_{S^2})^{-1}\propto \frac{1}{g_s \alpha' N} $. A very 
important point to notice here is that these KK modes are 
charged under $U(1)_T \times U(1)_R$ where the second 
$U(1)$ is a subgroup of the $SU(2)_R$ that is left 
untouched in the twisting procedure. On the other hand, the pure gauge
fields gluon and the gluino are not charged under 
either of the
$U(1)$'s.     

The dynamics of these KK modes mixes with the dynamics of 
confinement in this model because the strong coupling scale of 
the theory is of the order of the KK mass. One way to evade the mixing 
problem would be to work instead with the full string solution, namely
the world-sheet sigma model on this background (or in the 
S-dual NS5 background) which would give us control over the duality 
to all orders in $\alpha'$, hence we would be able to decouple the
dynamics of KK-modes from the gauge dynamics. This direction is
unfortunately not yet available.  

The dynamics of these KK modes have not been 
studied in much detail in the literature. In \cite{Gimon:2002nr} a very
interesting object---the {\em annulon}---was introduced. 
It is composed out of a condensate of many KK modes. 
More interesting studies on the annulons in this and related 
models---some being non-supersymmetric---are in \cite{Apreda:2003gs}. 

We would like to argue in this paper that the proposal 
of Lunin and Maldacena \cite{Lunin:2005jy} opens a new path to 
approach the KK mixing problem in a controlled way. 
Let us consider the torus of the $\b$ transformations as
$U(1)_T\times U(1)_R$---that is in fact the only torus in the
non-singular solution (\ref{metric}) and it is given by shifts along
$\vf$ and $\tilde{\vf}$. Then the proposal of LM implies
that,

{\em  The $\b$ deformation of (\ref{metric}) generates a dipole
deformation in the dual field the theory, but only in the KK sector
that is the only sector which is charged under $U(1)_T\times U(1)_R$.}       

We propose that the deformation does not affect the 4D field theory
modes namely the gluon and the gluino, encouraged by the fact that 
they are not charged under this symmetry, as explicitly shown in 
(\ref{twistboson}) and (\ref{twistfermion}).  

Let us sketch the general features of this
dipole field theory and then present an explicit argument showing an
example of how the dynamics of the KK sector of the theory is affected
under the dipole deformation. In particular, 
we would like to show that the masses of the KK modes 
are shifted under the dipole deformation. 

One can schematically write a Lagrangian for these fields as follows:
\beq
L= -Tr[\frac{1}{4} F_{\mu\nu}^2 + i\lambda D\lambda- (D_\mu\Phi_i)^2 
- (D_\mu\xi_k)^2 + \Psi 
(i D - M)\Psi + 
M_{KK}^2 (\xi_k^2 + \Phi_i^2) + V[\xi,\Phi, \Psi]]
\label{lagrangiantwisted}
\eeq
The potential typically contains the scalar potential for the bosons,
 Yukawa type interactions and more. 
This expression is schematic because of (at least) two reasons. First
of all, the potential presumably contains very
complicated interactions involving the KK and massless fields. 
Secondly, there is
mixing between the infinite tower of spherical harmonics that are obtained
by reduction on $S^2$ and $S^3$. (We were being schematic
in the definitions of (\ref{twistboson}) and
(\ref{twistfermion}). For example a precise designation for $\Phi$
should involve the spherical harmonic quantum numbers $(l,m)$ on
$S^2$.)      

Let us now discuss the effect of the $\b$ deformation in more
detail. The $U(1)_T$ corresponds to shift isometries along $\vf$ and  
$U(1)_R$ corresponds to shift along $\tilde{\vf}$ in
(\ref{metric}). From the D5 brane point of view, the former is a
dipole charge and the latter is a global phase on the 6D fields. 
Here we only focus on the $\g$ transformation because as we discussed in the
previous section, only the real part of the $\beta$ deformation gives rise to
a regular dual geometry. In this case, the prescription of Lunin and
Maldacena \cite{Lunin:2005jy} tells us to deform the product of 
two fields in the
{\em superpotential} as follows:
\be\lab{gammadeform} 
X[\vec{x}_6]Y[\vec{x}_6] \to e^{i\g \left(\hat{Q}_X \hat{L}_Y
    -\hat{Q}_Y \hat{L}_X\right)}
 X[\vec{x}_6]Y[\vec{x}_6],
\ee
where $X$ and $Y$ are either of the fields that appear in 
(\ref{lagrangiantwisted}) and $Q$ and $L$ denotes the charges of the
indicated fields under $U(1)_R$ and $U(1)_T$ respectively. Note that 
the action of $\hat{L}$ on a field that is charged under $U(1)_T$ is a
dipole deformation. 

As a simple exercise, let us investigate the implications of the
$\gamma$ deformation in (\ref{gammadeform}) for the masses of the KK
modes. Mass terms in (\ref{lagrangiantwisted}) are coming from the
quadratic terms in the superpotential. Let us consider  the
following term in original (undeformed) superpotential, 
\be\lab{spotund}
W \sim M_{KK}\;\Phi_0^+[\vec{x}_6]\;\xi_+^+[\vec{x}_6]\; +\; \cdots
\ee
Here we take two fields of the types in (\ref{twistboson}) with the
denoted charges under $U(1)_T$ and $U(1)_R$: $\Phi$ has charge +1 under
$U(1)_T$ and uncharged under $U(1)_R$. $\xi$ carries +1 under both. 
Under the deformation (\ref{gammadeform}) this term becomes,
\bea\lab{spgd}
  W&\to& W_{\g} \sim M_{KK}\,e^{i\g \left(\hat{Q}_{\Phi} \hat{L}_{\xi}
    -\hat{Q}_{\xi} \hat{L}_{\Phi}\right)}
\,\Phi_0^+[\vec{x}_6]\,\xi_+^+[\vec{x}_6]= M_{KK}\,
e^{-i\g R_3 R_2
  \hat{L}_{\Phi}}\,\Phi_0^+[\vec{x}_6]\,\xi_+^+[\vec{x}_6]\nn\\
{}&\approx& M_{KK}\,
\Phi_0^+[\vec{x}_6]\,(1+\g\,(R_2R_3)\,\6_{\vf})\,\xi_+^+[\vec{x}_6].
\eea
In the last step in (\ref{spgd}) we used the fact that the dipole
generator $\hat{L}$ corresponds to shifts in $\vf$

Here $R_2$ and $R_3$ are length scales that are associated with the
two-cycle and the three-cycle in the geometry.
\footnote{Note that $\g$ has
dimensions of $\a'^{-1}$ therefore in order to make the exponential
dimensionless the charges $Q$ and $L$ should have dimensions of 
length.}  
In general these length scales depend on
the radial coordinate $r$, hence the charges $Q$ and $L$ will depend on
the energy scale of the field theory. In the far IR region---that we
are ultimately interested in---they are both proportional to 
$\sqrt{\a'g_s N}$. Indeed, we support this claim by our supergravity
computations in section 5.3. 

Now, consider the spherical decomposition of $\xi^+_+$ on $S^2$:
\be\lab{sphdec}
\xi^+_+[\vec{x}_6] = \sum_{l,m}\xi^+_+[\vec{x}_4]Y_{l,m}[\q,\vf].
\ee
The spherical harmonics satisfy \footnote{for a detailed study of the KK modes 
spectrum see \cite{Andrews:2005cv}}, 
\be\lab{eiv}
 -i \6_{\vf} Y_{l,m}(\q,\vf) = m Y_{l,m}(\q,\vf).
\ee
Then, consider a specific KK mode of $\xi^+_+$ with the quantum numbers
$l,m$ in eq.(\ref{spgd}). Expanding (\ref{spgd}) for small $\g$ and
using (\ref{eiv}) we get the following particular term in the deformed
superpotential,
\be\lab{spotspfc}
W_{\g} \sim M_{KK} \Phi_0^+[\vec{x}_6]\xi_+^+[\vec{x}_4;l,m](1+i\g R_2R_3 m).
\ee
There is a mass term in the scalar potential of the theory that
derives from 
$$\bigg|\frac{\6 W_{\g}}{\6\Phi_0^+}\bigg|^2$$
and gives the following mass for this specific KK mode:
\be\lab{massdeform}
 (M_{KK}^{\g})^{2}  = \left(1+ (\g R_2R_3 m)^2\right) M^2_{KK}. 
\ee
 
Of course, our computation should be viewed as
schematic because in order to obtain the true mass eigenstates one
should diagonalize the mass matrix on infinite dimensional space of 
KK harmonics. Nevertheless, it is enough to support our main claim 
that 
the dipole deformation of the ${\cal N}=1$ theory changes 
the dynamics of the KK modes in a
very specific and controlled way. It is also enough to show that
the change in the masses of the KK modes are always in the positive
direction, they increase. We shall verify the mass
shift in (\ref{massdeform}) by explicitly computing the volumes of the
two and three cycles in the IR geometry in section 5.3. There we
obtain analytical expressions---as functions of $\g$--- 
for volume ratios of deformed over undeformed theories for the $S^2$ and $S^3$       
cycles. It is encouraging to see that the mass ratio is greater than
one. This fact is quite tempting to believe that the $\b$ deformation
of \cite{Maldacena:2000yy} is a positive step in curing the main
problem associated with this model and similar ones, 
namely mixing 
with the KK
modes. Our computations therefore 
suggest that for large values of
$\g$, one may be able to decouple the KK modes from the IR dynamics; 
but the value of $\g$ is to be restricted from above by the condition
of small Ricci scalar that is necessary for the validity of
supergravity approximation. Therefore we seem to have a finite window
for the $\gamma$ parameter where we can make an improvement of the
model by pushing the KK modes up. 

Taking very large values of $\gamma$ might 
conflict with having a bounded 
Ricci scalar to stay in the Supergravity approximation. 

Another encouraging fact is directly seen when one considers applying
the deformation in (\ref{gammadeform}) in case either of $X$ or $Y$ is 
the vector multiplet. As the vector multiplet is uncharged under
either of the $Q$ or $L$ we see that the $\gamma$ transformation of
the the ${\cal N}=1$ theory indeed does not affect the pure gauge
dynamics. It acts on the gauge and KK sectors of the theory (but it 
may translate into the massless sector through interaction terms). 
Motivated by the mass calculation that we presented
above, one can consider more elaborate computations regarding the
interaction terms in the potential in (\ref{lagrangiantwisted}). 
An interesting question to ask is how is the IR theory ``renormalized''
under the $\gamma$ transformations.   
        
\section{Deformations of the Singular ${\cal N}=1$ Theory}
\subsection{The Singular Solution $a(r)$=0}
In order to apply the $SL(2,R)$ transformation we shall put the metric
in (\ref{metric}) in the form explained in  \cite{Lunin:2005jy} 
(See also Appendix A in this paper). Like before, we will stick with 
the 
definitions
given in this paper to avoid confusion. 
It is very useful to first consider the
transformation of the simpler but singular case of $a(r)=0$. This 
provides
us with some intuition and  physical insight. 
The metric of eq.(\ref{metric}), written in string frame, 
in the case $a(r)$ set to zero 
reads as follows:
\beq
ds^2=F\left(\frac{1}{\sqrt{\Delta}}( d\varphi - C d\tilde{\varphi} 
+  {\cal A}^{(1)} - C {\cal 
A}^{(2)})^2 + 
\sqrt{\Delta}( 
d\tilde{\varphi}  +  {\cal A}^{(2)})^2\right) + 
\frac{e^{2/3\phi}}{F^{1/3}} g_{\mu\nu} 
dx^\mu dx^\nu
\label{metricform}
\eeq
where we write the eight dimensional part of the 
metric that is transverse to the
torus as, 
\beq
g_{\mu\nu}dx^\mu dx^\nu=  e^{-2/3\phi}F^{1/3} [D d\psi^2 + e^{\phi} 
(dx^2_{1,3}\,+\,g_s N\alpha'(e^{2h}d\theta^2+\,
dr^2 + \frac{1}{4} d\tilde{\theta}^2))],
\label{metricform2}
\eeq
we have defined the following vectors,  
\beq
{\cal A}^{(1)}= \alpha d\psi, \;\;{\cal A}^{(2)}= \beta d\psi,
\label{Avectors}
\eeq
and the  functions $F, D, \Delta, C, \alpha, 
\beta$ that appear above are:
\bea
& & F=\frac{\alpha' g_s N e^{\phi}}{4}
(4 e^{2h} \sin^2\theta+\cos^2\theta \sin^2\tilde{\theta} )^{\half}, 
\;
\Delta=\frac{4 e^{2h} \sin^2\theta + \sin^2\tilde{\theta} 
\cos^2\theta}
{(4 e^{2h} \sin^2\theta + \cos^2\theta)^2 }, 
\nonumber\\
& &  C = -\frac{\cos\theta \cos\tilde{\theta}}{4 e^{2h} 
\sin^2\theta +\cos^2\theta},D = \frac{\alpha' g_s N  e^{2h +\phi} 
\sin^2\theta
 \sin^2\tilde{\theta}}{4 e^{2h}\sin^2\theta + 
\sin^2\tilde{\theta}\cos^2\theta},
\nonumber\\
& & \alpha  = \frac{\cos\theta \sin^2\tilde{\theta}}{\cos^2\theta 
\sin^2\tilde{\theta} + 4 
e^{2h} \sin^2\theta}, \; 
\beta= \frac{4 e^{2h} \sin^2\theta cos\tilde{\theta}}
{ 4 e^{2h} \sin^2\theta + \cos^2\theta\sin^2\tilde{\theta} }
\eea
The RR two form given in (\ref{RR}) for $a(r)=0$ can be written as, 
\bea
C_2 &=& C_{12} (d\varphi +  {\cal A}^{(1)}) \wedge (d\tilde{\varphi} +  {\cal 
A}^{(2)}) 
+ C^{(1)}_\mu  (d\varphi +  {\cal A}^{(1)})\wedge dx^\mu + C^{(2)}_\mu  
(d\tilde{\varphi} +  
{\cal A}^{(2)})\wedge dx^\mu \nonumber\\
{}& & -\frac{1}{2}({\cal A}^{(a)}_\mu C^{(a)}_\nu - \tilde{c}_{\mu\nu}) dx^\mu 
\wedge dx^\nu.
\label{c2nos} 
\eea
Here various components and  one-forms read as follows:
\bea
& & C_{12}= -\frac{1}{4} \cos\theta \cos\tilde{\theta},\;\;C^{(1)}=- C_{12}\beta d\psi 
-\frac{\psi}{4} \sin\theta d\theta  ,\;\;
C^{(2)}= C_{12}\alpha d\psi + \frac{\psi}{4} \sin\tilde{\theta} d\tilde{\theta}, 
\nonumber\\
& & \tilde{c}_2 = \frac{\psi}{4}(\alpha \sin\theta d\psi \wedge d\tilde{\theta} +\beta 
\sin\tilde{\theta} d\tilde{\theta} \wedge d\psi)
\label{moredef}
\eea
Like in the previous section, various components of the two-form and 
the $A$-vectors above are
grouped in combinations that
transform under $SL(2,R)$ as vectors:  
\beq
V^{(i)}_\mu =(-\epsilon^{ij} B^{(j)}_\mu,A^{(i)}_\mu,\epsilon^{ij} 
C^{(j)}_\mu)= (0, A^{(i)}_\mu,\epsilon^{ij}C^{(j)}_\mu),\; 
W_{\mu\nu} =(\tilde{c}_{\mu\nu}, \tilde{d}_{\mu\nu},\tilde{b}_{\mu\nu})= 
(\tilde{c}_{\mu\nu},0,0)
\eeq
that transform explicitly as,
\bea
& & (V^{(i)})'= (- \epsilon^{ij} B^{(j)}, 
A^{(i)} + \gamma \epsilon^{ij} B^{(j)} - \sigma 
\epsilon^{ij} C^{(j)}, \epsilon^{ij} C^{(j)})\;\;W'= (\tilde{c} 
+\gamma \tilde{d}, \tilde{d}, \tilde{b} +\sigma \tilde{d})
\label{vector}
\eea
We will  concentrate on the $\gamma$-transformation. In this case, 
the vectors and tensors in (\ref{vectors}) transform as
$
(V^{(i)})'= (0, A^{(i)}, \epsilon^{ij} C^{(j)}),\;\; W'= (\tilde{c}, 
0, 0)$.
Notice that due to the transformations above, the differentials $D\varphi, 
D\tilde{\varphi}$ do not change under $SL(3,R)$.
Following what we explained in the previous section, we can 
transform the functions appearing in (\ref{metricform}).
We obtain the new values for $F, e^{\phi}, C_{12}, \chi, 
B_{12}$ as given by eq.(\ref{gentrans}).
For the $D5$ branes, 
the $\gamma$-transformation leads to nonsingular 
spaces according to the discussion of the previous section, we will 
concentrate in the following on transformations that 
have $\sigma=0$. Then the new fields become,
\beq
F'=  \frac{F } {(F^2\gamma^2 +1)},\;
B_{12}'=\frac{F^2 \gamma}{(F^2\gamma^2 +1)},\;
e^{2\phi'}= \frac{e^{2\phi} }{(F^2\gamma^2 +1)},\;\chi'=\gamma 
C_{12}, \;
C_{12}'=\frac{C_{12}}{(F^2\gamma^2 +1)}. 
\label{functions}
\eeq
Putting all of the things together, we can get the new 
configuration, that consists in  this case of  metric, dilaton, 
axion, RR and NS two forms and 
four form.
The new metric, in string frame is given by, 
\bea
& & ds^2= \frac{F'}{\sqrt{\Delta}} (D\varphi - C D\tilde{\varphi})^2 
+ 
F'\sqrt{\Delta}(D\tilde{\varphi})^2 +(\frac{e^{2 
\phi'}}{F'} e^{-2\phi} 
F)^{1/3} 
[D d\psi^2 +\nonumber\\
& & e^{\phi} (dx^2_{1,3}\,+\,\alpha' g_s 
N(e^{2h}d\theta^2+ dr^2 + \frac{1}{4} 
d\tilde{\theta}^2))]
\label{metrictransformed}
\eea
the NS gauge field is
$B_{\varphi,\tilde{\varphi}} = B'_{12} D\varphi' \wedge 
D\tilde{\varphi'}$, with $B_{12}'$ given in (\ref{functions}).
The new dilaton and axion are  
$e^{2\phi'}= \frac{e^{2\phi}}{(1+ \gamma^2 F^2)}, \;\;\chi'=
 -\frac{\gamma}{4}\cos\theta \cos\tilde{\theta}$,
the new RR two-form,
\eqn{newRR2}{ C^{(2)'} = C'_{12} D\varphi \wedge
D\tilde{\varphi} -  C^{(1)} \wedge D\varphi -  C^{(2)} \wedge 
D\tilde{\varphi} -\frac{1}{2} (A^{(1)}\wedge C^{(1)} + 
A^{(2)}\wedge C^{(2)} 
-\tilde{c} ),} with $C_{12}'$ given in (\ref{functions}). Finally, 
the RR four form is 
given by
\beq
(C_4)'= -\frac{1}{2}B_{12}' (\tilde{c} - (A^{(1)}\wedge C^{(1)} + A^{(2)}\wedge C^{(2)}) 
) \wedge D\varphi \wedge D\tilde{\varphi}.
\label{fourform}
\eeq
For completeness, let us write here the expressions for the gauge 
field strengths $H_3', F_3', F_5'$
\beq
H_3'= dB_{12}' \wedge (d\varphi \wedge d\tilde{\varphi} +\alpha 
d\psi \wedge d\tilde{\varphi} +\beta d\varphi \wedge d\psi) + 
B_{12}'(d\tilde{\varphi}\wedge d\alpha + d\beta \wedge 
d\varphi)\wedge d\psi
\label{h3p}
\eeq
\bea
& & F_3'= dC_2 -\chi' H_3'=\nonumber\\
& &  \frac{1}{4} d\psi \wedge 
(\sin\theta 
d\theta \wedge d\varphi - \sin\tilde{\theta} d\tilde{\theta} \wedge 
d\tilde{\varphi}) +
d\varphi  \wedge d\tilde{\varphi} \wedge \Big(\frac{1}{1+\gamma^2 
F^2} dC_{12} + C_{12} d(\frac{1-\gamma^2 F^2}{1+\gamma^2 F^2})\Big) 
+ \nonumber\\
& & (-\alpha d\tilde{\varphi} +\beta d\varphi  )\wedge 
d\psi \wedge 
\frac{\gamma^2 F^2}{1+\gamma^2 F^2} dC_{12}
\label{F3p}
\eea
and the RR four form that reads
\beq
-2 C_4'= B_{12}'( \tilde{c}- {\cal A}^{(i)}\wedge C^{(i)})\wedge 
D\varphi \wedge D\tilde{\varphi}
\eeq
and finally
\beq
F_5'= dC_4' - C_2'\wedge H_3'
\eeq
What can we learn from this transformed background?
One first thing that comes to mind is related to chiral symmetry 
breaking. Indeed, as shown in \cite{Gursoy:2003hf}, it is enough to 
work with the singular background and the respective RR fields to 
see explicitly the phenomena of $\chi$SB as a Higgs mechanism for a 
gauge field used to gauge the isometry corresponding to chiral 
symmetry (translations in the angle $\psi$ in this case).
It should be interesting to do again this computation in this 
transformed background. The new ingredients to take into account are 
the new NS and RR fields as shown above.
We believe that the anomaly will be the same because, as an 
anomaly, it can only be affected by the mass less fields. In our case, 
the transformed background only takes into account changes in the 
dynamics of the massive KK modes. It should be interesting to see 
this argument working explicitly.
\subsection{Transformations in the R-symmetry Directions}
In the previous section, 
we performed rotations that commuted with the 
$R$ symmetry of the field theory. 
Indeed, the $R$-symmetry of N=1 SYM is 
represented in the background studied in 
in the previous subsection by changes in the angle 
$\psi$. The rotations done above, 
preserve SUSY, since they do not 
involve the angle $\psi$. 
In this section, we will concentrate on rotations
taking the torus, to be composed of $\psi, 
\tilde{\varphi}$; this will break SUSY.
Notice that in this case, the dual field theory 
to the transformed solution 
will not be a dipole theory, but a 
theory where phases has been added to the
interaction terms. 
Also, notice that we 
can do this in the singular 
solution only, because in the desingularized solution, 
we do not have the invariance $\psi\to 
\psi + \epsilon$.

So, to remind the reader, let us  write explicitly the 
10 metric (\ref{metric}) in the singular case (string frame 
is used)
\bea
& & ds^2_{10}\,=\,\alpha' g_s N e^{{\phi}}\,\,\Big[\,
\frac{1}{\alpha' g_s N }dx^2_{1,3}\,+\,
e^{2h}\,\big(\,d\theta^2+\sin^2\theta
d\varphi^2\,\big)\,+\,
dr^2\,+\nonumber\\
& & \,{1\over 4}\,(d\tilde{\theta}^2 + \sin^2\tilde{\theta} d\tilde{\varphi}^2 
+(d\psi + 
\cos\theta d\varphi +  \cos\tilde{\theta} 
d\tilde{\varphi} )^2) \,\Big]\,\,,
\eea
We notice that this metric is already written in 
the form that we want it.
Indeed, comparing we can see that
\beq
F=\frac{\alpha' g_s N}{4} e^{{\phi}} \sin\tilde{\theta}, 
\;\;\Delta= \sin^2\tilde{\theta}, 
\;\;D\varphi_1= D\psi= d\psi + \cos\theta d\varphi, 
\;\; D\varphi_2 =D\tilde{\varphi}= 
d\tilde{\varphi}, \; C=- \cos\tilde{\theta}.
\label{defr}
\eeq
and 
\beq
g_{\mu\nu} dx^\mu dx^\nu
= e^{-2\phi/3}F^{1/3}[\alpha' g_s N e^{{\phi}}  \big(\,
\frac{1}{\alpha' g_s N }dx^2_{1,3}\,+\,e^{2h}\,\big(\,d\theta^2+\sin^2\theta
d\varphi^2\,\big)\,+\,
dr^2\, + \frac{d\tilde{\theta}^2}{4} \big)]
\eeq
After doing some computations, see Appendix D, 
we can see that the transformed string metric reads,
\bea
& & (ds^2_{string})'\,=\, e^{2(\phi'-\phi)/3}(\frac{F}{F'})^{1/3}\alpha' g_s N 
e^{{\phi}}\,\,\Big[\,
\frac{1}{\alpha' g_s N }dx^2_{1,3}\,+\,e^{2h}\,\big(\,d\theta^2+\sin^2\theta
d\varphi^2\,\big)\,+\,
dr^2 \nonumber\\
& & +{1\over 4}\, d\tilde{\theta}^2\Big]+
(F' \sqrt{\Delta} (D\tilde{\varphi})^2 
+\frac{F'}{\sqrt{\Delta}}(D \psi +  \cos\tilde{\theta} 
D\tilde{\varphi} )^2) 
\label{transformedRsym}\eea

The new RR and NS fields are explicitly written in the Appendix D.
The reader  can check that  there is no 
five form generated. Indeed, there is 
no $C_4$ generated by the SL(3,R) rotation
and besides the term $C_2 \wedge H_3$ does not contribute.

This new solution is expected to be non-SUSY, and an occasional 
resolution of 
the 
singularity at $r=0$ could give a dual to 
YM theory. One might think about, for example, a resolution by 
turning on a 
black hole, dual to YM at finite temperature. This should be very 
interesting to solve.

Like in the examples of the D5 brane, doing this transformation
is changing quantities like 
for example the Ricci scalar, but the place where 
$\alpha' R_{eff}$ diverges 
($r\to 0$ with a divergence of the form $\alpha' 
R_{eff}\approx r^{-7/4}$) are the same before and after 
the transformation.
The divergent structure does not seem to 
become worst by the effect of the  gamma 
transformation.

\subsection{The General $\b$ Transformation}

Transformations of the singular solution \ie (\ref{metric}) for $a=0$
where the torus is chosen transverse to the $D5$ brane have the 
interesting property of being regular {\em both for real and
  imaginary parts of $\b$}. Therefore it is tempting to perform the
general $\b=\gamma+i\sigma$ transformation. Here we, show the
regularity of the general transformed solution explicitly, by
presenting the resulting geometry.    

The only new feature that one introduces to the results of
section 4.2 is that, turning on $\s$ changes the connections 
${\cal  A}^{(i)}$ according to (\ref{vectors}). The transformed
connections are: 
\beq
{\cal A}^{(1)'} = {\cal A}^{(1)} = \cos\tt d\vp,\;\;
{\cal A}^{(2)'}={\cal A}^{(2)} + \sigma C^{(1)} = \frac{\a'N}{4}\s\cos\q\vf,   
\eeq
Details are given in Appendix D. 
Using (\ref{gentrans}), we find the the string frame metric as, 
\bea
& & ds^2_{string}\,=\, 
U_{st} \alpha' 
g_s N 
e^{{\phi}}\,\,\Big[\,
\frac{1}{\alpha' g_s N }dx^2_{1,3} +
e^{2h}\big(\,d\theta^2+\sin^2\theta
d\varphi^2\big)+
dr^2 +{1\over 4} d\tilde{\theta}^2\Big] \\
& & + F' \sqrt{\Delta} (d\tilde{\varphi}+\s\frac{\a'N}{4}\cos\q d\vf)^2 
+\frac{F'}{\sqrt{\Delta}}(d\psi + \cos\theta 
d\varphi +\cos\tilde{\theta} 
d\tilde{\varphi} + \sigma\frac{\a'N}{4}\cos\tt\cos\theta d\varphi )^2)
\nonumber\\
\eea
Here the new metric functions are given by (\ref{gentrans}) and
(\ref{ust}) and they generally depend on both $\gamma$ and $\sigma$.
Eqs. (\ref{gentrans}) also determine how the RR two-form and the
dilaton transforms and shows the new fields that are generated 
by the general transformation.  
   
It should be interesting to study the fate of the symmetry 
$\psi\to\psi+\epsilon$ after this transformation.
This symmetry, in the case of $N=1$ SYM
was associated with chiral rotations \cite{Gursoy:2003hf}. In this 
case, our theory does not have massless fermions. So, the symmetry 
corresponds to some kind of flavor symmetry within the KK sector. 
It should be interesting to see if this symmetry gets 
by the broken and the way studied in \cite{Gursoy:2003hf}.

Now that we gained some insight with this type of configurations, 
let us study the $U(1)\times U(1)$ transformation in the case of the 
non-singular background. 
\section{Deformation of the Non-singular ${\cal N}=1$ Theory}
Now, let us discuss the non-singular case, $a(r)\ne 0$. The type of 
problems we have in mind that might be tackled, are related to 
previous computations of non-perturbative field theory aspects from 
a  
supergravity perspective. Indeed, in some cases, it was not clear if 
the result of this computation was afflicted by the presence of the 
massive KK modes. So, finding the transformed background and 
re-doing the computations with it, might help us improve this 
situation, since as we remarked above, both backgrounds do differ in 
the fact that their 
dual theories have different dynamics for the KK modes, results that 
depend on the transformation parameter $\gamma$ will indicate the 
presence of effects of the KK modes.
 
We first write the full metric 
in (\ref{metric})
in the following appropriate form for the $SL(3,R)$ rotation in which 
the torus and the transverse parts have been separated as follows:

\bea
& & ds_{string}^2= e^{\phi} [ dx_{1,3}^2 + \alpha' g_sN dr^2] + 
D_1 d\psi^2 + D_2 d\theta^2 + D_3 d\tilde{\theta}^2 +  
E_1 d\theta d\tilde{\theta}+
 \nonumber\\
& & E_2 d\theta d\psi  + E_3  d\tilde{\theta} d\psi 
+\frac{F}{\sqrt{\Delta}} 
[d\varphi + (\alpha_1 - C\beta_1) d\theta + (\alpha_2 - C\beta_2) 
d\tilde{\theta} +(\alpha_3 - C\beta_3) d\psi - C d\tilde{\varphi}]^2+\nonumber\\
& & F\sqrt{\Delta} (d\tilde{\varphi} 
+ \beta_1 d\theta + \beta_2 d\tilde{\theta} + \beta_3 d\psi)^2 
\label{newmetrica}
\eea

To simplify the notation let us define,
\be\label{S}
f =  4e^{2h}\sin^2\theta+\cos^2\theta+a^2\sin^2\theta, \;\; g= a \sin\theta
\sin\tilde{\theta} \cos\psi-\cos\theta \cos\tilde{\theta}.
\ee
Then, various functions in (\ref{newmetrica}) are given as,
\bea
& & F = \frac{\alpha' g_sN e^{\phi}}{4} \sqrt{f-g^2},\;\; \Delta =
\frac{f-g^2}{f^2},\;\; C = \frac{g}{f},\;\; \nonumber\\
& & \beta_1 =  \frac{f}{f-g^2}
a \sin\psi \sin\tilde\theta,\;\; \beta_2 = \frac{g}{f-g^2} a \sin\psi
\sin\theta,
\;\; \beta_3  = \frac{f \cos\tilde\theta + g \cos\theta}{f-g^2}\nonumber\\ 
& &  
\alpha_1= \frac{a \sin\tilde{\theta} \sin\psi g}{f-g^2},\;\; \alpha_2= \frac{a
\sin\theta \sin\psi}{f-g^2}, \;\; \alpha_3 = \frac{\cos\theta + g
\cos\tilde{\theta}}{f-g^2}\nonumber\\
& & D_1=\frac{\alpha' g_sN e^{\phi} }{4(f-g^2)}(f
\sin^2\tilde{\theta}-g^2-\cos^2\theta-2g \cos\theta
\cos\tilde{\theta}),\;\;\nonumber\\
& & D_2= \frac{\alpha' g_sN e^{\phi}}{4} (a^2+4e^{2h}-\frac{f}{f-g^2}a^2\sin^2\psi
\sin^2\tilde\theta)\\
& & D_3 = \frac{\alpha' g_sN e^{\phi}}{4} (1-\frac{a^2\sin^2\theta
  \sin^2\psi}{f-g^2}),\;\; E_1= \frac{a}{2}  \alpha' g_sN e^{\phi }  (\cos\psi -
\frac{g}{f-g^2}a \sin^2\psi \sin\theta \sin\tilde{\theta})\nonumber\\
& & E_2 = -\frac{ \alpha' g_sN e^{\phi } a}{2} \frac{\sin\psi \sin\tilde\theta (f
\cos\tilde\theta + g \cos\theta)}{f-g^2}, \;\; 
E_3  = -\frac{ \alpha' g_sN e^{\phi }}{2}  \frac{a \sin\theta \sin\psi}{f-g^2} (\cos\theta + g
\cos\tilde{\theta})\nonumber.
\label{definitionnonsing}
\eea
Now, let us focus on the RR two form.
Like before, it is useful to define four one forms as 
${\cal A}^{(i)}, {\cal C}^{(i)}, \; 
i=1,2$
\beq\label{defne1}
{\cal A}^{(1)}=\alpha_1 d\theta +\alpha_2  d\tilde{\theta} + \alpha_3 d\psi, \;\;
{\cal A}^{(2)}= \beta_1 d\theta + \beta_1 d\tilde{\theta} + \beta_3 d\psi, \;\;
\eeq
and 
\beq\lab{defne2}
{\cal C}^{(1)}=C^{(1)}_{\theta} d\theta +C^{(1)}_{\tilde{\theta}}   d\tilde{\theta} 
+C^{(1)}_{\psi}   d\psi, 
\;\;
{\cal C}^{(2)}=C^{(2)}_{\theta}  d\theta + C^{(2)}_{\tilde{\theta}}  d\tilde{\theta} 
+C^{(2)}_{\psi}   d\psi, 
\;\;
\eeq
and the two form $\tilde{c}_{\mu\nu}dx^\mu \wedge dx^\nu$, with components
$\tilde{c}_{\theta\tilde{\theta}}, \tilde{c}_{\tilde{\theta} \psi},$ and 
all others being zero.
In order to reproduce the RR two-form potential given in 
(\ref{RR}) in the following form, 
\eqn{newRR2p}{ C^{(2)} = C_{12} D\varphi \wedge
D\tilde{\varphi} -  C^{(1)} \wedge D\varphi -  C^{(2)} \wedge 
D\tilde{\varphi} -\frac{1}{2} (A^{(1)}\wedge C^{(1)} + A^{(2)}\wedge C^{(2)} 
-\tilde{c} );}
the quantities defined above are determined as,
\bea
& & C_{12}= \frac{1}{4} (a(r) \sin\theta \sin\tilde{\theta} \cos\psi - \cos\theta 
\cos\tilde{\theta}),\;\;\nonumber\\ & & 
{\cal C}^{(1)}=(-\frac{\psi}{4} \sin\theta - 
C_{12}\beta_1) d\theta -(\frac14 a(r) \sin\theta \sin\psi +
C_{12}\beta_2)d\tilde{\theta} -\beta_3 
C_{12} d\psi , \;\; \nonumber\\
& & {\cal C}^{(2)}= (\frac14 a(r) \sin\tilde{\theta} \sin\psi +
C_{12}\alpha_1)d\theta +(\frac{\psi}{4} \sin\tilde{\theta} 
-C_{12}\alpha_2)d\tilde{\theta} + \alpha_3 C_{12} d\psi, 
\;\;\nonumber\\
& & \tilde{c}= \tilde{c}_{\theta \tilde{\theta}} d\theta \wedge 
d\tilde{\theta} +  \tilde{c}_{\theta \psi} d\theta \wedge d\psi +
\tilde{c}_{\tilde{\theta} \psi} d\tilde{\theta} \wedge d\psi 
\eea
with
\bea
& & \tilde{c}_{\theta \tilde{\theta}}=
\alpha_1\alpha_2+\beta_1\beta_2+C_{12}(\alpha_1\beta_2-\alpha_1\beta_2)
-\frac14 \psi \sin\tilde{\theta}\beta_1 +\frac14 a \sin\theta \sin\psi
\alpha_1\nonumber\\
& & -\frac14 a \cos\psi + \frac14 a \beta_2 \sin\psi \sin\tilde{\theta}, \;\; \nonumber\\
& & \tilde{c}_{\theta
  \psi}=\alpha_1\alpha_3+\beta_1\beta_3-C_{12}(\alpha_1\beta_3-\alpha_3\beta_1)
-\frac14 \psi \sin\theta \alpha_3+\frac14 a \sin\psi \sin\tilde{\theta}
\beta_3 , \nonumber\\
& & \tilde{c}_{\tilde{\theta}
  \psi} = \alpha_2\alpha_3+\beta_2\beta_3-C_{12}(\alpha_2\beta_3-\alpha_3\beta_2)
+\frac14 \psi \sin\tilde{\theta} \beta_3-\frac14 a \sin\psi \sin\theta \alpha_3 .
\label{definitionsnoab}
\eea
One can check that in the limit $a(r)\to 0$ one recovers the result in
(\ref{metricform}).

So, after the $SL(3,R)$ rotation we obtain a metric
that reads
\bea
& & (ds_{string}^2)'=(\frac{e^{2(\phi'-\phi)} F}{F'})^{1/3}
\Big( e^{\phi} [ dx_{1,3}^2 + 
\alpha' 
g_sN dr^2] + 
 D_1 d\psi^2 + D_2 d\theta^2 + D_3 d\tilde{\theta}^2 +  
E_1 d\theta d\tilde{\theta} +\nonumber\\
& &  E_2 d\theta d\psi  + E_3  
d\tilde{\theta} d\psi \Big)
+\frac{F'}{\sqrt{\Delta}} 
[d\varphi + (\alpha_1 - C\beta_1) d\theta + (\alpha_2 - C\beta_2) 
d\tilde{\theta} +(\alpha_3 - C\beta_3) d\psi - C d\tilde{\varphi}]^2\nonumber\\
& & + F'\sqrt{\Delta} (d\tilde{\varphi} 
+ \beta_1 d\theta + \beta_2 d\tilde{\theta} + \beta_3 d\psi)^2 
\label{newmetrica2}
\eea
The RR and NS fields will transform according to the rules discussed 
in previous sections. Notice that as happened before, the $\gamma$ 
transformation leaves the one forms
${\cal A}^{(i)}, \; C^{(i)}$ invariant. 
This is a
good point to notice that the torus given by  the cycle of constant 
$\theta, 
\tt, \psi, r,x,$ has volume $V_{T2}= F'= 
\frac{F}{1+\gamma^2 F^2}$ after the 
transformation and that only vanished if $F$ vanishes or becomes 
infinite. So, if the 
original geometry is nonsingular, the transformed one also is. 
In the points where the volume of the torus shrinks, 
it happens that it does in a way $B_{12}'=Re[\tau]\to 0$
when $\sqrt{g'}=Im[\tau]\to 0$, thus the metric satisfies the 
criteria in \cite{Lunin:2005jy} for nonsingularities.

\subsection{Confinement}

The first check that this new solution should pass is the
confinement. As clearly explained for example in 
\cite{Sonnenschein:1999if}, the expectation value of the Wilson loop 
can be computed by calculating the Nambu-Goto action for a string  
that is connected to a probe brane at infinity and explores the IR region 
(r=0) of the background. The criteria for confinement is that if
the following combination (that can be intuitively associated with the tension of 
the confining string) 
\beq
T_s=\sqrt{g_{tt} g_{xx}}\bigg|_{r=0}
\eeq
is non-vanishing then there is linear confinement. The ``QCD string'' tension, 
$T_s$ defined 
above, is non vanishing.
In the original background, 
the value of $T_s= e^{\phi(0)}$. After the transformation, 
we can see that the string tension is given by
\beq
T_{s}'= \sqrt{g_{tt}' g_{xx}'}= U_{st} \sqrt{g_{tt} g_{xx}}= T_s
\eeq
where $U_{st}$ is defined in (\ref{ust}). 
We note that the tension would change if we 
perform a $\sigma $-transformation. It should be interesting to 
understand the effects of the sigma transformation in more detail, 
since they seem to alter important aspects of the gauge theory. 

\subsection{The Beta Function} Let us compute the
beta function, following the lines of  
\cite{DiVecchia:2002ks}, \cite{Bertolini:2002yr}. 
We first revise  their steps and add 
some comments that might make the derivation more clear. 
In
order to compute beta functions, 
one needs to define a SUSY cycle where to wrap a $D5$
brane at finite distance from the origin. It was found in 
\cite{Nunez:2003cf} that such cycle exists for an
infinite value of the radial constant. 
The cycle is given by the following identifications 
\footnote{notice that also the cycle $\theta=\pi- 
\tilde{\theta}, \;\; \varphi = 
\tilde{\varphi}, \;\;\psi=0,2  \pi$ 
is SUSY at large values of $r$} 
\beq
\theta=\tilde{\theta}, \;\; 
\varphi = 2\pi - \tilde{\varphi},\;\;\psi=\pi, 3\pi 
\label{cycle} 
\eeq
Notice that, since this is a 
calibrated (SUSY) cycle for large values of the radial coordinate, we
should be considering the 
``abelian version'' ($a(r)=0$) of the background. We proceed with
the full nonsingular solution, but 
one should keep in mind that this is a good approximation
only for $r\to \infty$. Then, 
one needs to introduce a relation between the radial coordinate r, and
the energy scale of the theory.
We will take the following relation \cite{Apreda:2001qb},
\cite{DiVecchia:2002ks}, that identifies
the gaugino condensate with the function $a(r)$.
\beq
a(r)=\frac{\Lambda^3}{\mu^3}=\langle\lambda\lambda\rangle
\label{id}
\eeq 
This relation has an intuitive explanation since 
the gaugino condensate, like the turning-on of the function 
$a(r)$ are phenomena that occur in the IR, that is at small values 
of the energy/radial coordinate.
This identification (\ref{id}), is indeed the reason 
for doing the 
calculation in the non-singular background, even 
when the cycle we use is supersymmetric only for 
large values of $r$. 

Using (\ref{cycle}) one defines 
the coupling constant as
\beq
\frac{1}{g_{YM}^2}=\frac{1}{(2\pi)^2 
g_s\alpha'}\int_{0}^{2\pi}d\varphi 
\int_{0}^\pi d\theta e^{-\phi} (g^{xx})^2 \sqrt{det G_6}.
\label{coup}
\eeq
This gives,
\beq
\frac{\pi^2}{4 g_{YM}^2 N}= (4 e^{2h} +(a(r)^2-1)^2)
\label{coup2}
\eeq
Therefore on obtains the beta function as
\beq
\beta= \frac{dg_{YM}}{dlog(\mu/\Lambda)}= -\frac{g_{YM}^3 N}{8 \pi^2} 
\frac{d(\frac{\pi}{4g^2N})}{dr} (\frac{d(-1/3 
Log(\mu/\Lambda))}{dr})^{-1}.
\label{betaa}\eeq
Expanding the result for large values of the radial coordinate
and using an expansion for large $r$ of (\ref{coup2}) one gets,
\beq
\beta= -\frac{3 g_{YM}^3 N}{8\pi}\frac{1}{(1- 
\frac{g^2_{YM}N}{8\pi})}
\eeq
If one keeps higher orders in the $r$ expansion, one gets extra 
terms that in 
\cite{DiVecchia:2002ks} were attributed to fractional instantons.

The reader may wonder what is the origin of these extra effects.
On one hand, one might think that they are a pure 
${\cal N}=1$ SYM effects. Or might argue that they are an effects due to 
the KK modes of the field theory to which this supergravity solution 
is dual. Apart from these two possibilities, one can think
that they are spurious effects coming from 
the fact that for small values of $r$ in the expansion 
of the quantities above, the supersymmetry is broken, since the cycle 
(\ref{cycle}) is no longer SUSY.

In order to discard some of these alternatives,
we compute the beta function in the $\g$ deformed 
non-singular background. If these extra terms that seem to modify the 
original NSVZ result are effects of the KK modes, then by our general
philosophy in this paper, the beta function should be different in 
the deformed theory. 

Let us take the same cycle in the transformed solution 
(\ref{newmetrica2})
\footnote{ We believe that by general properties of the $\g$ transformation
this cycle is also supersymmetric in the deformed geometry.  
It should be interesting to verify this by an explicit BPS
calculation.}
The relevant six dimensional part of the metric is
\bea
& & (ds_{6D}^2)'= [ e^{\phi} (dx_{1,3}^2 + 
g_s N \alpha' dr^2 ) + (D_2+ D_3 + E_1)d\theta^2 ]+ 
\nonumber\\
& & \frac{F'}{\sqrt{\Delta}}( (1+C) d\varphi + (\alpha_2 - C\beta_2) 
d\theta)^2 +
F'\sqrt{\Delta})(-d\varphi+ (\beta_1+\beta_2)d\theta)^2
\label{metricacycle}
\eea
We note that at the special cycle the metric components reduce to the
original undeformed components except than the change from $F$ to
$F'$. This gives a factor of $F'/F$ which exactly cancels out the
change in the dilaton in (\ref{coup}). 
So, computing the determinant and defining the coupling as before,
\beq
\frac{1}{g_{YM}^2}= 
\frac{1}{(2\pi)^2 g_s\alpha'}\int_{0}^{2\pi} d\varphi \int_0^\pi 
d\theta e^{-\phi'}(g'^{xx})^2 \sqrt{det G_6} .
\label{couplingcon}\eeq 
Using the fact that
\beq
e^{\phi-\phi'}=(1+ \gamma^2 F^2)^{1/2}
\label{diladilap}
\eeq
the coupling reads
\beq
\frac{\pi^2}{4g_{YM}^2 N}= [4 e^{2h} + 
(a^2-1)^2].
\eeq
This is precisely the same as (\ref{coup2}). Then, one repeats the 
procedure in eq. (\ref{betaa})  
\cite{DiVecchia:2002ks}, 
\cite{Bertolini:2002yr} and obtains the same result for the beta
function in the deformed theory.  

One can repeat the same computation for the theta parameter of
Yang-Mills. The essentials of the computation do not change.
$\theta_{YM}$ is the same as in the undeformed theory.

We learn that the computation for beta function 
seems robust under the deformations. Indeed,  the beta function 
is a field theory result, that should be independent of the KK modes dynamics. 
So, this transformed solution that only changes the KK sector of the 
dual field theory, does not produce any effect in the result 
(\ref{betaa}).
On the other hand, the fact that we got the 
same result is perhaps indicating that the result 
should only be taken to first order in the large $r$ 
expansion  thus implying that there are no
``fractional instanton'' corrections.
Or may be, the ``fractional instantons'' are a genuine field theory
effects; but the result of these computations above, make sure that 
the KK modes have no relation to them whatsoever.
Finally, we note that it should of course be of great interest to find the SUSY cycle 
that computes the beta function in the IR.
\subsection{KK Modes and the Domain Wall Tension}
There are two types of KK modes, those that are proportional to the 
volume of an $S^2$ in the geometry and those that are proportional 
to the volume of a three cycle, the first type are the `gauge' KK 
modes. On the other hand,
the volume of the non-vanishing three cycle at the IR (at 
$r=0$) is inversely related to the masses of the ``geometric'' KK 
modes in the
theory. Therefore it is very interesting to see whether or not
these volumes are changed by the transformation. If the volume 
decreases
under the transformation, it would be a non-trivial improvement of 
the
model as the undesired KK degrees of freedom would 
have better decoupling. Let us start with the `geometric' KK modes. 
The three cycle in the original theory is given by, 
\beq
\theta=\varphi=r=\vec{x}=0.
\label{ckk}
\eeq
In the original metric (\ref{metric}), this volume is, 
(in the string frame), 
\be\lab{mkk}
V= 2\pi^2 (\alpha' g_sN e^{\phi(0)})^{\frac32} .
\ee
Now, we consider the same cycle in
the transformed geometry (\ref{newmetrica2}). Using
eqs. (\ref{definitionnonsing}) the reader can see that one obtains
\bea\lab{masskks3}
Vol(S^{3'}) &=& \pi^2 (\alpha' g_s N e^{\phi_0})^{3/2} 
\int_{0}^{\pi} d\tt 
\frac{\sin\tt}{1+ \mu^2 \sin^2\tt}\nonumber\\
{}& = & \pi^2 (\alpha' g_s N e^{\phi_0})^{3/2} 
(\frac{2 \atanh(\mu/\sqrt{1+\mu^2})}{\mu \sqrt{1+\mu^2}})\nonumber\\
{}&\propto & \pi^2 (\alpha' g_s N e^{\phi_0})^{3/2} 
(2 -\frac{4}{3}\mu^2 +..),\nonumber
\eea
where $16\mu^2=\gamma^2 (\alpha' g_s N e^{\phi_0})^{2}$.  
It is useful to consider the ratio,
\beq
\left(\frac{M_{KK}'}{M_{KK}}\right)^2= \left(\frac{Vol S^3}{Vol S^{3'}}\right)^{2/3}= 
\left(\frac{\mu \sqrt{1+\mu^2}}{2 \atanh[\mu/\sqrt{1+\mu^2}]}\right)^{2/3} .
\eeq
So, we see that the mass of these `geometric' KK modes indeed 
increase improving the decoupling between the KK and the pure SYM sector.  

Now, let us analyze the mass of the `gauge' KK modes, that are 
inversely proportional to the volume of some $S^2$ defined in the 
geometry. Let us define the two cycle as,
\beq
\tt=\pt=\psi=x=r=0, (\theta,\varphi)
\eeq
Computing the line element of this two cycle one gets (once 
again, we concentrate on the $\gamma$ transformation)
\beq
ds^2= (D_2 +\beta_1^2 F'\sqrt{\Delta})d\theta^2 + 
\frac{F'}{\sqrt{\Delta}}d\varphi^2
\eeq
so, the volume of this two cycle near $r=0$ is
\bea
& & Vol(S^{2'})= \frac{\pi \alpha' g_s N 
e^{\phi_0}}{2}\int_{0}^{\pi} 
\frac{d\theta}{\sqrt{1+\frac{1}{16}(\gamma \alpha' g_s N e^{\phi_0} 
\sin\theta)^2}}\to \nonumber\\
& &  Vol(S^{2'})\approx \frac{\pi \alpha' g_s N e^{\phi_0}}{2}(1 - 
\frac{\gamma^2 \alpha'( g_s N)^2 e^{2\phi_0}}{64}) 
\eea
again, it is convenient to present the quotient,
\beq
(\frac{m'_{kk}}{m_{kk}})^2= (\int_{0}^{\pi}
\frac{d\theta}{\sqrt{1+\frac{1}{16}(\gamma \alpha' g_s N e^{\phi_0}
\sin\theta)^2}})^{-1}\approx 1 + \frac{1}{16}(\gamma \alpha' 
g_s N e^{\phi_0})^2
\eeq
So, we see that the mass of the `gauge' KK modes also increases 
when this transformation is performed.
We would like to stress again  that 
under the $\s$ transformation, the volumes above  indeed
change. However, as the $\s$ transformation produces irregularities, 
the role of the $\s$ modification of the original
model is unclear \footnote{ We are grateful to Juan Maldacena for suggesting
 that redefinitions of the periodicities of the original cycles may
 improve the situation.}     

Another interesting quantity to compute is the tension of a Domain 
Wall. A domain wall is thought as a $D5$ brane that wraps the three 
cycle (\ref{ckk}) and extends in $(2+1)$ directions of spacetime.
The way of computing the tension of this wall is by computing the 
coefficient in front 
of the Born-Infeld action for a $D5$ as indicated above.
\beq
S_{wall}= \int d^3 x (\int d\Omega_3 
e^{-\phi'}\sqrt{det G_6}) + CS
\eeq
from the term in parentheses we read the tension of the domain wall, 
to be
\beq
T_{wall}= \int_{0}^{4\pi}\int_{0}^{2\pi} \int_{0}^{\pi}  
d\tilde{\theta} 
d\tilde{\varphi} d\psi e^{-\phi' +2\phi} (e^{2(\phi'-\phi)} 
\frac{F}{F'}\alpha' g_s N)^{1/2} \frac{F}{(1+ \gamma^2 
F^2)}\bigg|_{r=0}
\eeq
Using (\ref{diladilap}) we see that the result changes and the chane
is explicitly given as, 
\be\lab{dwallogrh}
\frac{T'}{T}= \half+\frac{1}{\pi}\arctan(\frac{4e^{-\phi_0}}{\g\a'g_s
  N }).
\ee  

One should perhaps interpret this as 
an effect of the KK modes in field theory 
observables because the cycle we are choosing (\ref{ckk})
is not the appropriate one for computing domain walls. This point 
deserves more study.

In this line, it would be nice to analyze
what happens to the computation of baryonic strings as D3 branes 
wrapping the previous 3 cycle done in the paper
\cite{Hartnoll:2004yr}. 

The result of the last two subsections 
is basically that under the $\gamma$ transformations
many field theory aspects of N=1 SYM are 
not changed when computed from the
transformed background, while the KK modes change their masses. 
This, on one hand is 
reassuring of the conjectured duality, because 
as we stressed many times, the transformed 
background only differs in the 
dynamics of the KK modes. It is 
of interest to see an example where some 
changes in the dynamics 
are indeed expected to happen. This example 
is provided by 
pp-waves. The objects called annulons and studied in 
\cite{Gimon:2002nr},
\cite{Apreda:2003gs}  are constructed to be heavy composites 
of many 
KK modes (this composite is called `hadron' in the papers mentioned 
above). These objects are studied in a plane wave approximations 
of the geometry. So, in order to see real changes in dynamics of KK 
modes, in the next section,we will study these composites in the 
transformed backgrounds.

\section{PP-waves of the Transformed Solutions}

PP-wave limits of various transformed solutions are quite
interesting. The first example of this kind was discussed by Lunin and
Maldacena \cite{Lunin:2005jy} where they considered a particular PP-wave limit
of the $\b$ transformed $AdS_5\times S^5$ geometry; it is quite 
amusing to observe that with different arguments and objectives 
Niarchos and Prezas found the same plane wave some years ago
\cite{Niarchos:2002fc}. In most cases, 
one can indeed obtain PP-waves that lead to quantizable string
actions. Here, we would like to first make some observations about
the general properties that are satisfied by the PP-waves of the
transformed 
geometries. Then, we exemplify these properties by studying two simple cases, 
those of flat the $D5$ brane and singular $D5$ wrapped on $S^2$. Finally
we discuss the PP-wave associated to the transformed non-singular
geometry that we obtained in section 4.2.

We restrict our attention to the pp-wave limits that are obtained by
the scaling of the {\em same} coordinates, before and after the transformation. 
The motivation for this is to study the effects of the $SL(3,R)$ 
transformation to the pp-wave limits. In general one may also ask
  whether or not the transformed geometry admits interesting pp-wave
  limits other than the pp-waves of the original geometry. A separate
  issue is to apply the $SL(3,R)$ transformation to the pp-wave
  itself and seek for new quantizable geometries. 
We note that generally the pp-wave limit does not commute with the 
$SL(3,R)$ transformation. We do not investigate these interesting
issues further in this paper.        

\subsection{General Properties}

For all of the geometries that we consider, 
the original metric is proportional to $e^{\phi}$ (in the string frame) and the
metric before the
$SL(3,R)$ transformation can generally be written in the following
form:
\be\lab{pp1}
ds^2 = \frac{F}{\sqrt{\Delta}}\left(D\vf_1-C D\vf_2\right)^2+
F\sqrt{\Delta}(D\vf_2')^2 + e^{\phi}\left( -dt^2 + d\vec{x}^2 + p(r) dr^2 + q(r)
d\Omega^2\right)
\ee 
where $r$ is a radial coordinate $p$ and $q$ are some given
functions and  $\Omega$ is a compact space. It is important to notice
that, the function $F$ is proportional to $e^{\phi}$. We allow for singular
geometries, because it is well-known that singularities are usually
smoothed out in the pp-wave limit.  

Suppose that
the radial coordinate runs from $r_0$ to infinity. We define
$e^{\phi(r_0)} = R^2$. A general class of pp-waves are obtained 
by scaling some of the coordinates as follows:
$$ r\to r_0 + \frac{r}{R},\; \vec{x}\to \frac{\vec{x}}{R},\;
\phi_i\to\frac{\phi_i}{R}, \; R\to\infty, $$
where $\phi_i$ are some angular coordinates on $\Omega$ or the along torus. 

Since $F\sim e^{\phi}$, one observes that, in this limit
$$F \sim R^2 f[\frac{r}{R},\frac{\phi_i}{R},\cdots]$$
where $f[\dots]$ is a function of the indicated variables. We assume
that this function behaves like ${\cal O}(1)$ or ${\cal
  O}(R^{-1})$.  Therefore the function $F$ blows up in
the limit as $ F\sim R^2\; {\rm or}\; F\sim R. $ 
Also, in many cases, $\sqrt{\Delta}\sim {\cal O}(1)$ or ${\cal O} (R^{-1})$ in this limit. 
Thus, we have,
\be\lab{pw}
F\sim R^t,\;t\in\{1,2\},\qquad \sqrt{\Delta}\sim R^{-p},\;p\in\{0,1\}.
\ee

Now, consider applying the same
pp-wave limit to the geometry after the transformation, \ie\, to the new
geometry that is given by (\ref{newmetric}). We limit our attention only to $\gamma$
transformations for simplicity. In this case, $U_{st}=1$ and $F$ in 
(\ref{pp1}) is replaced by,
$$F'  = \frac{F}{1+\gamma^2 F^2}.$$
From the above scaling behaviour of the function $F$, 
 we see that the torus in the transformed geometry shrinks to zero
 size, hence yields a singular pp-wave limit {\em unless one also
   scales $\gamma$ appropriately.}
Consider the following scaling:
$$ \gamma R^s = \tilde{\gamma} =const.,\; s>0$$     

If in addition to regularity of the limit, one asks for a linear
pp-wave that has the potential of being quantizable, one would also
like the denominator in $F'$ be expanded in powers of
$\tilde{\gamma}$. Then, a glance at (\ref{pp1}) and (\ref{pw}) leads us to consider
the following, most appropriate scaling:
\be\lab{sfix}
s = \half(3t+p).
\ee
Let us denote the pp-wave limit of the original geometry by
$ds^2_{pp}$ and the same limit after the $SL(3,R)$ transformation by 
$ds^{'2}_{pp}.$ Then, if we fix $s$ as above, in the $R\to\infty$
limit we obtain the following general result:
\be\lab{ppgen}
ds^{'2}_{pp}= ds^2_{pp} - \tilde{\gamma}^2 ds^2_{pp-torus}.
\ee
Here the second term is the pp-wave limit of only the torus part of
the geometry. 
There is no guarantee that (\ref{sfix}) will allow for a linear
pp-wave, however this is ---in some sense--- the best one can do in
order to avoid
non-linearity in the pp-wave. In addition---as we assume
that the original metric has a nice, linear pp-wave
limit---we are able to isolate the non-linearity --if any-- in the
second term in (\ref{ppgen}).   

\subsection{PP-wave of the Transformed Flat $D5$}
 
Let us consider the flat $D5$ geometry, (\ref{D5sol}), as our first
example. A smooth linear pp-wave limit can be obtained by performing
the following coordinate transformations, (we define $\eta^2 = \a' g_s N$
\beq
r\to r_0 +\frac{r}{R},\;\;\; \vec{x}_5 \to\frac{\vec{x}_5}{R},\;\;\; 
\tt=\frac{\tt}{R}
\label{scalingd5}
\eeq
We also define
\beq
dt= dx^+ +\frac{dx^-}{L^2},\;\;\; d\psi +d\pt=\frac{2}{\eta} (dx^+ 
-\frac{dx^-}{L^2}),\;\;\; e^{\phi(r_0)}=R^2.
\eeq

The limit $R\to\infty$ yields the following geometry,
\beq\lab{ppd5}
ds^2_{pp1}= - 4dx^+ dx^- +d\vec{x}_5^2  + \eta^2dr^2 
+ \frac{\eta^2}{4}(d\tt^2 +\tt^2 d\pt^2) -  
\frac{\eta^2}{4} \tt^2 dx^+ d\pt. 
\eeq

Now let us consider applying the same pp-wave limit,
(\ref{scalingd5}), to the case of the transformed
$D5$ brane that we discussed in section 3.3. We have the 
geometry, given by (\ref{newvecs}) and (\ref{newD5}). We consider the
regular case, \ie\, $\sigma=0$. In this case we see from
(\ref{metfuncD5}) that 
\be\lab{fdd5}
 F \to \frac{\eta^2}{4} R \tt,\;\;\; \sqrt{\Delta}=\frac{\tt}{R}.
\ee 
Therefore $t=p=1$ in (\ref{pw}). From (\ref{sfix}) we see that the
appropriate scaling of $\gamma$ as $s= 2$ \ie \, $\tilde{\gamma}=\gamma
R^2 = fixed$. Then from (\ref{ppgen}) one obtains the pp-wave that
results from the transformed D5 solution as follows:
 \beq\lab{pptD5}
ds^{'2}_{pp1}= ds^{2}_{pp1} - \tilde{\gamma}^2 (\frac{\eta^2}{4})^3 \tt^2 dx^{+2},
\eeq 
where $ds^{2}_{pp1}$ is given in (\ref{ppd5}). 
 We see that the $SL(3,R)$ transformation produced a simple $\gamma$-
 correction in the original pp-wave geometry. This fact was 
 observed in \cite{Lunin:2005jy} in case of the deformed $AdS_5\times
 S^5$. The string theory is quadratic and easily 
quantizable and one observes that the
 bosonic $\tt$ field receives a correction to its mass that is proportional to
 $\tilde{\gamma}^2$. 

\subsection{PP-wave of the Transformed D5 on S2}

Let us now apply our general discussion to a slightly more complicated
example: The singular geometry of $D5$ brane wrapping an $S^2$. The
metric is, 
\beq
ds^2= e^{\phi}\Big[ dx_{1,3}^2 +\eta^2 \Big(dr^2  + r 
(d\theta^2 + \sin^2\theta d\varphi^2) +\frac{1}{4}( d\tt^2 + 
\sin^2\tt d\pt^2 + (d\psi + \cos\theta d\varphi + \cos\tt 
d\pt)^2)\Big)  \Big]
\label{d5ons2singular}
\eeq
This geometry is supplied with a dilaton $e^{2\phi} 
=\frac{e^{2\phi_0 + 2r}}{\sqrt{r}}$ and an RR 
three form. We take a similar geodesic as the one above (\ref{scalingd5}):
\beq
r\to r_0 +\frac{r}{R}, \;\; \theta\to\frac{\theta}{R},\;\; 
\tt\to\frac{\tt}{R}, \;\; \vec{x}_3\to \frac{\vec{x}_3}{R},\; dt= dx^+ 
+\frac{dx^-}{R^2}, \;\; d\psi + d\varphi +d\pt = 
\frac{2}{\eta}(dx^+ -\frac{dx^-}{R^2}) 
\label{scalingd5s2}
\eeq
where $R$ is again defined by $e^{\phi(r_0)}= R^2$. The following
linear pp-wave geometry follows from the $R\to\infty$ limit:
\beq
ds_{pp2}^2=  -4 dx^+ dx^-+ d\vec{x}_3^2 +\eta^2 dr^2 +\eta^2 r_0 
(d\theta^2 +\theta^2 d\varphi^2) +\frac{\eta^2}{4}(d\tt^2 +\tt^2 
d\pt^2) - (\theta^2 d\varphi +\tt^2 d\pt)dx^+  
\label{d5s2singpp}
\eeq

Now we consider the $\gamma$ transformation where we take the same torus as
is section 5.2. $F$ and $\sqrt{\Delta}$ behaves exactly the same as in
(\ref{fdd5}). Therefore, we also fix $s=2$ and we get the following new
pp-wave from (\ref{ppgen}) by focusing on the same geodesic as
in (\ref{scalingd5s2}):

\beq
ds_{pp2}^{'2}= ds_{pp2}^2 - \tilde{\gamma}^2 (\frac{\eta^2}{4})^3 \tt^2dx^{+2} 
\label{pptd5s2}
\eeq
where $ds_{pp2}^2$ is given in (\ref{d5s2singpp}). 

\subsection{PP-wave Limit of the Transformed Non-singular Solution}

We finally consider the physically most interesting case of the
transformed non-singular solution. It has proved somewhat tricky to
obtain a regular pp-wave limit of this solution even before the
$SL(3,R)$ transformation. This limit is discussed in
\cite{Gimon:2002nr}. Here we first review the argument 
of \cite{Gimon:2002nr} and we revise it slightly so that it can be
applied also for the case of $\g$ deformed non-singular background. 

In order to explore the gauge dynamics at IR one is interested 
in a geodesic near $r=0$. With this purpose in mind, the 
authors of \cite{Gimon:2002nr} considered the
 non-singular ${\cal N}=1$ geometry in (\ref{metric})
and picked up a null geodesic that is on the $S^3$ at the origin. 
It is tricky to find the suitable coordinates for this geodesic
essentially because of the following fact: the one-forms $w^i$ of
(\ref{su2}) that involve the angular coordinates of $S^3$ are fibered
by the $S^2$ coordinates and this fibration is given by the connection
one form $A^i$. As one scales $r\to r/R$ and 
the angles on $S^3$ by $\phi_i\to \phi/R$ together, one does
not obtain a metric that is suitable for the pp-wave limit. This is  
because the the one-forms are ${\cal O}(1)$ near $r\sim 0$ (the
function $a(r)$ approaches to 1 as $r\to 0$). Therefore this part of
the metric blows up as $R\to\infty$. 

This only indicates that one
should use a better set of coordinates that is more suitable for the
pp-wave limit of this geometry. \cite{Gimon:2002nr} solved this
problem as follows. The geometry when Scherk-Schwarz reduced on the
$S^3$ produces an $SU(2)$ gauged supergravity in 7D
\cite{Chamseddine:1997nm}. On the other hand it is well-known that
the gauge field $A$---that was the cause of the problem that we
mentioned above--- is pure gauge at the origin, up to ${\cal O}(r^2)$
corrections:
\be\lab{puregauge}
A=-i dh h^{-1} + {\cal O}(r^2),\;\;\; h=
e^{-i\s^1\frac{\q}{2}}e^{-i\s^3\frac{\vf}{2}}.
\ee
Here, we defined $A=A^i\frac{\s^i}{2}.$
Therefore \cite{Gimon:2002nr} simply gauged the ${\cal O}(1)$ part of
$A$ away by the following gauge transformation,
\be\lab{gaugetr}
A\to h^{1}A\,h + ih^{-1} d h,
\ee
and then took the appropriate pp-wave 
limit as 
\beq
r\to\frac{r}{R}, \;\;  
\tt\to\frac{\tt}{R}, \;\; \vec{x}_3= \frac{\vec{x}_3}{R},\; dt= dx^+ 
+\frac{dx^-}{R^2}, \;\; d\psi +d\pt = 
\frac{2}{\eta}(dx^+ -\frac{dx^-}{R^2}). 
\label{nonsinglim}
\eeq
This limit produces the linear, quantizable pp-wave that is given in
\cite{Gimon:2002nr}. 
Here, we would like to consider the same pp-wave limit that is given in 
(\ref{nonsinglim}) in the case of the $\g$ deformed solution
(\ref{newmetrica2}). As the $S^3$ of the undeformed solution is
distorted by the $\g$ transformation, one cannot simply make a ``gauge
transformation'' in $A$. Below, we explain the appropriate way 
to put the metric in a suitable form.   

The gauge transformation (\ref{gaugetr}) is nothing else but a coordinate 
transformation on the angular variables $\tt,\pt,\psi$ 
from the 10D point of view. Explicitly, it is equivalent to changing
the coordinates as,
\be\lab{coordtr}
w\to h\,w\,h^{-1}-idh\,h^{-1},
 \ee
where we defined $w=w^i\frac{\s^i}{2}$. Noting that the one-forms,
(\ref{su2}) on the $SU(2)$ group manifold are given as follows,
$$ w = -i dg\, g^{-1},\;\;\; 
g=e^{i\s^3\frac{\psi}{2}}e^{i\s^1\frac{\tt}{2}}e^{i\s^3\frac{\tvf}{2}}$$
one sees that the transformation in (\ref{coordtr}) is equivalent to 
\be\lab{coordtr2}
   g\to h\,g.
\ee
where $h$ is given in (\ref{puregauge}). We give the explicit form of
this coordinate transformation in Appendix E. 
Now we apply this
coordinate transformation to the $\g$ deformed metric in
(\ref{newmetrica2}). Using the explicit expressions in Appendix E 
one can work out the full coordinate transformation of
(\ref{newmetrica2}). However, it is sufficient for us to observe the
following. At the end of section 6.1 we argued that, 
if one knows the pp-wave limit in the undeformed solution
one can obtain the new pp-wave of the deformed solution 
from (\ref{ppgen}). Having performed the coordinate transformation  
(\ref{coordtr2}), we put the first term in (\ref{ppgen}) in the
 appropriate form to perform the pp-wave limit in
 (\ref{nonsinglim}). Therefore the limit of this part will be
 explicitly given as the pp-wave in \cite{Gimon:2002nr}, that is
\beq
ds^2= -2 dx^+ dx^- - \frac{1}{g_s N \alpha'}(\frac{1}{9}\vec{u}^2 + \vec{v}^2)(dx^{+})^2 + 
d\vec{x_3}^2 + dz^2 + d\vec{u}^2 + d\vec{v}^2
\label{ppwaveleo}
\eeq
Here $u$ and $v$ are two-planes and $z$ is a line. 
There is also an expression for the three form.
The second
part in (\ref{ppgen}), then produces some non-linear corrections to
this pp-wave that is proportional to $\tilde{\g}$. In order to obtain
a meaningful limit we found that the scaling should be defined as
follows:
$$ \gamma R^4 = \tilde{\gamma} =const.$$ 

We explicitly see that the $\g$ deformation produces a non-linear
correction to the pp-wave of the original solution and this
deformation should be dual to the complicated dynamics that affects
the KK-sector of the ${\cal N}=1$ theory.

\section{Deformations of the ${\cal N}=2$ Theory}

With the fields ($A_\mu^a, \phi^a, \psi^a$) mentioned
in section 2, one can perform a different twisting from the 
one explained there. One can choose the second $U(1)$
to twist, inside the diagonal combination of the $SU(2)'s$. That is 
$U(1)_D$ inside diag($SU(2)_L \times SU(2)_R$). If this is done, we 
can see that the massless spectrum can be put in correspondence with 
a vector multiplet of ${\cal N}=2$ in d=(3+1).
The corresponding solution, preserves eight supercharges as was 
found in 
\cite{Gauntlett:2001ps}

\bea\lab{N2metric}
  ds^2 &=& e^{\f} \left(dx_{1,3}^2+\frac{z}{\l}\left(d\tt^2+\sin^2\tt d\pt^2\right)\right)+\frac{e^{-\f}}{\l}\left(d\r^2+\r^2d\vf_2^2\right)\\
{} & & + \frac{e^{-\f}}{\l z}\left( d\tilde{\s}^2+\tilde{\s}^2\left(d\vf_1+\cos\tt
    d\pt\right)^2\right),
\eea
where $\l = (g_s N \a')^{-1}$. 
The dilaton is,
\be\lab{N2dilaton}
 e^{2\f} = e^{2z}\left(1-\sin^2\q\frac{1+c e^{-2z}}{2z}\right).
\ee
Here $z$ and $\q$ are given in terms of the radial functions that
appear in (\ref{N2metric}) as 
\be\lab{N2cov}
\r=\sin\q e^z,\;\; \tilde{\s}=\sqrt{z}\cos\q e^{z-x}
\ee
and $x$ is the following radial function:
\be\lab{N2x}
e^{-2x} = 1-\frac{1+c e^{-2z}}{2z}.
\ee
$c$ is an integration constant. 
The RR two-form field reads, 
\be\lab{N2RR}
C^{(2)} = g_s N\a' \vf_2 d\left( \xi (d\vf_1 + \cos\tt d \pt)\right),
\ee
where
\be\lab{ksi}
\xi = (1+e^{2x}\cot^2\q)^{-1}.
\ee

It is well-known that the R-symmetry of the theory is dual to the
isometry along the $\vf_2$ direction. This chiral symmetry is 
anomalous
because of the $\vf_2$ dependence of $C^{(2)}$.  

\subsection{Transformation along a non-R-symmetry Direction} 

We shall first discuss the simpler case of $SL(2,R)$ transformations along
the torus that is composed of $\vf_1$ and $\pt$. The metric is already
in the desired form given in eq. (\ref{metricform}) 
\be\lab{N2expmetrpp}
ds^2=F\left(\frac{1}{\sqrt{\Delta}}( d\vf_1 - C d\pt +  {\cal A}^{(1)} - C {\cal 
A}^{(2)})^2 + 
\sqrt{\Delta}( 
d\pt  +  {\cal A}^{(2)})^2\right) + g_{\mu\nu} dx^\mu dx^\nu
\ee
with,
\be\lab{N2nonR}
F=\frac{\tilde{\s} \sin\tt}{\l},\; \Delta = \frac{e^{2\phi}z^2
  \sin^2\tt}{\tilde{\s}^2},\; C = - \cos\tt,\; {\cal A}^{(1)} = {\cal A}^{(2)} = 0
\ee
and 
\be\lab{intmetp}
g_{\m\n}dx^{\m}dx^{\n} = e^{\f}
  (dx_{1.3}^2+\frac{z}{\l}d\tt^2)
+ \frac{e^{-\f}}{\l z}d\tilde{\s}^2 + \frac{e^{-\f}}{\l}(d\r^2+\r^2d \vf_2^2). 
\ee

The RR two from is put in the following form:
\bea
C_2 &=& C_{12} (d\vf_1 +  {\cal A}^{(1)}) \wedge (d\tilde{\varphi} +  {\cal 
A}^{(2)}) 
+ C^{(1)}_\mu  (d\vf_1 +  {\cal A}^{(1)})\wedge dx^\mu + C^{(2)}_\mu  
(d\tilde{\varphi} +  
{\cal A}^{(2)})\wedge dx^\mu \nonumber\\
{}& & -\frac{1}{2}({\cal A}^{(a)}_\mu C^{(a)}_\nu - \tilde{c}_{\mu\nu}) dx^\mu 
\wedge dx^\nu.
\label{N21c2nos} 
\eea
Comparison of (\ref{N2RR}) with (\ref{N21c2nos}) yields the following
components of the RR two-form:
\bea\lab{N2RRcomps}
C_{12} & = & 0,\; C^{(1)}_{\r}=-\frac{\vf_2}{\l}\frac{\6\xi}{\6\r},\;
C^{(1)}_{\tilde{\s}} = -\frac{\vf_2}{\l}\frac{\6\xi}{\6\tilde{\s}},\;\nonumber\\
C^{(2)}_{\r} & = & -\frac{\vf_2}{\l}\frac{\6\xi}{\6\r}\cos\tt,\; 
C^{(2)}_{\tilde{\s}}=-\frac{\vf_2}{\l}\frac{\6\xi}{\6\tilde{\s}}\cos\tt,\; 
C^{(2)}_{\tt}=\frac{\vf_2}{\l}\xi \sin\tt,\; \tilde{c}_{\m\n}=0. 
\eea

The transformation yields the following results. 
\be\lab{N2nonRtrans}
{\cal A'}^{(1)} = -\tilde{\s} C^{(2)},\; {\cal A'}^{(2)} = \tilde{\s} C^{(1)}, \;
{C'}^{(1)} = C^{(1)},\; {C'}^{(2)} = C^{(2)},\; {\tilde{c}'}_{\m\n}=0.
\ee

New metric functions are given by the general formula,
eq. (\ref{gentrans}) with $C_{12}=0$. In the simpler case of $\s=0$
transformation one obtains, 
\be\lab{simptrans}  
C'_{12} = 0, \; F'=\frac{F}{1+\g^2F^2},\; 
B'_{12} = \frac{\g F^2}{1+\g^2F^2}, \; e^{2\f'} =
\frac{e^{2\f}}{1+\g^2F^2},\; \chi'=0. 
\ee

Now, let us see whether or not the gravity dual computations of the
$\b$-function and the chiral anomaly are affected by the $SL(2,R)$
transformation. We shall consider the general transformation that
depends on both $\s$ and $\gamma$ in what follows. 
Gravity duals of the theta
angle and the YM coupling constant can be obtained by placing a probe
D5-brane on the supersymmetric cycle $\tilde{\s}=0$ in the geometry 
\cite{DiVecchia:2002ks}. From the DBI action on the probe brane one
can read off the information after reducing on the $S^2$. 
Essentially, the coupling
constant is determined by the volume of $S^2$ that the D5 branes
wrap and the theta angle is determined by the flux of the RR form on
this sphere. One gets the following results
(before the transformation): 
\be\lab{N2cc}
\frac{1}{g^2} = \frac{1}{2(2\pi)^3g_s \a'}\int_0^{2\pi}d
\pt\int_0^{\pi} d\tt e^{-\phi} Vol(S^2)= \frac{N}{4\pi^2}\log\r.
\ee
\be\lab{N2th}  
\theta_{YM}=\frac{1}{2\pi g_s\a'}\int_0^{2\pi}d
\pt\int_0^{\pi} d\tt C^{(2)}_{\tt\pt} = -2N\vf_2. 
\ee

We consider the transformed solution determined by the transformation
rules in (\ref{gentrans}) and (\ref{N2nonRtrans}). One obtains the
following result for the coupling constant:
\be\lab{N2cctrans}
\frac{1}{{g'}^2} = \frac{1}{2(2\pi)^3g_s \a'}\int_0^{2\pi}d
\pt\int_0^{\pi} d\tt e^{-\phi'} Vol(S^2)'= \frac{1}{2(2\pi)^3g_s \a'}\int_0^{2\pi}d
\pt\int_0^{\pi} d\tt H^{-\half} e^{-\phi} Vol(S^2).
\ee
Here $Vol(S^2)'$ denotes the volume of $S^2$ in the transformed
geometry and the function $H$ is given by, 
\be\lab{H}
H = 1+ \s^2 \frac{\tilde{\s}^2}{\l} \sin^2\tt e^{-2\f}.
\ee
However one should remember to place the probe brane on the
supersymmetric cycle. From the general transformation rules under
$SL(2,R)$ it is not hard to see that $\tilde{\s}=0$ maps to
$\tilde{\s}=0$ hence we have the same supersymmetric cycle. On this
cycle one learns from (\ref{H}) that $H=1$. Therefore we refer that
the coupling function is unaffected by the transformation. 

For the theta angle, one has the following new expression:
\be\lab{N2thtrans}  
\theta'_{YM}=\frac{1}{2\pi g_s\a'}\int_0^{2\pi}d
\pt\int_0^{\pi} d\tt {C^{(2)}}'_{\tt\pt} = 
\frac{1}{2\pi g_s\a'}\int_0^{2\pi}d 
\pt\int_0^{\pi} d\tt (1-JG\s)C^{(2)}_{\tt\pt}.
\ee
From (\ref{GHJ}) we see that $J=0$ at the supersymmetric cycle
$\tilde{\s}=0$ (Remember that when $C_{12}=0$). Therefore we learn that
$\theta_{YM}$ is also unaffected by the transformation.  
This result is indeed expected because the
particular $SL(3,R)$ transformation that we make here is not breaking
the supersymmetry. Then, this result can be viewed as a consistency check
for the whole setup.  

\subsection{Transformation along the R-symmetry Directions}

Now, let us work out a more complicated case of the transformation
involving the torus on the $\vf_2 $ and $\pt$ angles. As $\vf_2$ was
identified with the R-symmery we expect that this transformation
breaks supersymmetry. We rewrite the metric again in the form
(\ref{metricform}) that makes 
the torus explicit.
\be\lab{N2expmetrp}
ds^2=F\left(\frac{1}{\sqrt{\Delta}}( d\vf_2 - C d\tilde{\pt} +  {\cal A}^{(1)} - C {\cal 
A}^{(2)})^2 + 
\sqrt{\Delta}( 
d\pt  +  {\cal A}^{(2)})^2\right) + \frac{e^{2/3\phi}}{F^{1/3}} g_{\mu\nu} 
dx^\mu dx^\nu
\ee
with,
\be\lab{intmet}
g_{\m\n}dx^{\m}dx^{\n} = e^{-2\f/3}F^{1/3} \left( e^{\f}
  (dx_{1.3}^2+\frac{z}{\l}d\tt^2)
+ \frac{e^{-\f}}{\l z}d\tilde{\s}^2 + \frac{e^{-\f}}{\l}d\r^2+D
d\vf_1^2\right ). 
\ee
Various functions that appear in the metric is determined as,
\bea\lab{N2varfunc}
\frac{F}{\sqrt{\Delta}} & = & e^{-\f}\frac{\tilde{\s}^2}{\l},\; 
F\sqrt{\Delta}=e^{\f}\sin^2\tt\frac{z}{\l}+e^{-\f}\cos^2\tt\frac{\tilde{\s}^2}{z\l},\nonumber\\
D & = & e^{-\f}\frac{\tilde{\s}^2}{z\l} - F\sqrt{\Delta}{{\cal
A}_{\vf_1}^{(2)}}^2,\; {\cal A}^{(2)} =
e^{-\vf}\frac{\tilde{\s}^2}{z\l}\frac{\cos\tt}{F\sqrt{\Delta}} d\vf_1,\\
{\cal A}^{(1)} & = & C=0.
\eea
The RR two-form is written, 
\bea
C_2 &=& C_{12} (d\vf_2 +  {\cal A}^{(1)}) \wedge (d\tilde{\varphi} +  {\cal 
A}^{(2)}) 
+ C^{(1)}_\mu  (d\vf_2 +  {\cal A}^{(1)})\wedge dx^\mu + C^{(2)}_\mu  
(d\tilde{\varphi} +  
{\cal A}^{(2)})\wedge dx^\mu \nonumber\\
{}& & -\frac{1}{2}({\cal A}^{(a)}_\mu C^{(a)}_\nu - \tilde{c}_{\mu\nu}) dx^\mu 
\wedge dx^\nu.
\label{N22c2nos} 
\eea
Comparison of (\ref{N2RR}) with (\ref{N22c2nos}) yields the following
components of the RR two-form:
\bea\lab{N2RRcomps2}
C_{12} & = & 0,\; C^{(1)}_{\m}= 0, 
C^{(2)}_{\r} =  -\frac{\vf_2}{\l}\frac{\6\xi}{\6\r}\cos\tt,\nonumber\\  
C^{(2)}_{\tilde{\s}} & = & -\frac{\vf_2}{\l}\frac{\6\xi}{\6\tilde{\s}}\cos\tt,\; 
C^{(2)}_{\tt}=\frac{\vf_2}{\l}\xi \sin\tt,\; \tilde{c}_{\m\n}=0, \; 
\tilde{c}_{\vf_1 \tilde{\s}} =
\frac{\vf_2}{\l}\frac{\6\xi}{\6\tilde{\s}}(\half \cos\tt-1),\nonumber\\
\tilde{c}_{\vf_1 \r} & = & 
\frac{\vf_2}{\l}\frac{\6\xi}{\6\r}(\half \cos\tt-1),\;  
\tilde{c}_{\vf_1 \tt} = - 
\frac{\vf_2}{\l} \xi \sin\tt.\nonumber
\eea 

The transformation yields the following results. 
\be\lab{N22nonRtrans}
{\cal A'}^{(1)} = -\tilde{\s} C^{(2)},\; {\cal A'}^{(2)} = {\cal A}^{(2)}, \;
{C'}^{(1)} = 0,\; {C'}^{(2)} = C^{(2)},\; {\tilde{c}'}_{\m\n}={\tilde{c}}_{\m\n}.
\ee
New metric functions are given by the general formula,
eq. (\ref{gentrans}) with $C_{12}=0$. 

For the new coupling constant and the new theta angle it is not hard
to see that one would get exactly the same results had one allowed to
put the probe brane at the would-be susy cycle $\tilde{\s}=0$. However
as the transformation breaks supersymmetry, $\tilde{\s}=0$ is not a
supersymmetric cycle anymore. Therefore one does not trust the
computations of the beta function and the theta angle in this deformed
solution. It will be interesting to study these issues further. 

\section{Summary and Conclusions }

In this paper we studied in detail the proposal of Lunin and 
Maldacena\cite{Lunin:2005jy}, that 
is to interpret particular $SL(3,R)$ transformations of supergravity 
backgrounds as duals to field theories where the product between 
fields have been modified in a particular way.

We considered the quantum theories that correspond to 
the $\b$ deformations of D5 branes and analyzed the 
regularity of the transformed configuration.  

More importantly, we applied this to models that are argued to be 
dual to the strong coupling regime of ${\cal N}=1,2$ SYM.
In this case, $\beta$ transformations affect the so called 
KK modes of the theory, that can be thought as `contaminating' 
the pure SYM theory in the supergravity approximation. 
We argued by a simple
dipole field theory computation that the $\g$ deformations increase
the mass of the KK modes. However they do not directly affect the pure
gauge dynamics. We confirmed these expectations from the
supergravity side by computing  
some field theory observables in the transformed supergravity 
backgrounds. When these quantities were purely of 
field theory origin, no modifications were observed, thus providing 
a check for the original proposal.
We also showed that, as a consequence of the transformations, 
there is an increase in the mass of the KK modes, hence the 
decoupling is improved.
We would also like to stress here the fact that
in these transformed geometries, the dilaton does not blow at 
infinity. This fact was already observed in the non-commutative case 
and here we extend it for all $\beta$ 
deformations.

We have studied pp-waves of the configurations that are mentioned above. In 
this case, these pp waves are dual to some condensates of KK modes 
called annulons. We have studied changes in the dynamics of these 
annulons.

As a check of consistency, we also computed the dual to $N=1$ non-commutative SYM 
with the methods explained here and obtained the solution previously studied by 
\cite{Mateos:2002rx}. This is in Appendix B. 

Let us now comment on possible further work. It should be 
interesting to study the physical effects of these transformations 
on observables of the field theory. For example, by considering 
Wilson loops that wrap 
the transformed directions one could get some information that 
depends on the deformation parameters. Also, the study of 
strings rotating in the internal space could yield additional information.
Another field theory quantity that is interesting to study, and that is 
not expected to change, is the law obeyed by the tensions of confining 
strings. These models usually obey a sine law behaviour \cite{Herzog:2001fq}.

We did not check explicitly 
the SUSY of the deformed backgrounds, 
but we expect that, unless where we indicated, the same spinors will 
be preserved. In this line, it should be interesting to study the 
SUSY holomorphic two cycles to place probe D5 branes in the deformed 
geometry. This is interesting for the quenched SQCD and meson dynamics 
as studied in \cite{Nunez:2003cf}. It is possible that some of the 
holomorphic two cycles found in \cite{Nunez:2003cf} do indeed 
change.
These changes should be reflected in the dynamics and spectrum of mesons.

It will also be interesting to study the glueballs 
of the 
transformed models. Indeed, since
one of the problems of studying glueballs from a supergravity 
perspective is that we cannot disentangle those glueballs that 
belong to the original field theory from those that are coming from 
KK modes, the study of this spectrum for the transformed models is 
interesting because in the comparison we might be able to 
identify what belongs to the field theory we are interested in and 
what is an artifact of the supergravity approximation.
In this line, one can study the possibility of 
finding massless glueballs, following what is done in 
\cite{Gubser:2004qj}. In this case, the asymptotic behavior 
of the dilaton in the transformed solution is different, thus it might be 
possible 
to find a massless glueball. However, this should 
be analyzed in detail.
A study of the transformed  family of metrics developed in 
\cite{Butti:2004pk} may also be interesting. 

We also note that a more detailed study of the annulons in these 
confining models is required. The annulons are precisely made of 
the part of the spectrum that is modified by the $SL(3,R)$ 
transformations.

It should also be of interest to study the $G_2$ holonomy manifolds 
in M theory. These backgrounds can be thought of as coming from D6 
branes wrapping a three manifold inside a $CY3-$fold. There are at 
least three U(1)'s that can be identified 
(see \cite{Hartnoll:2002th}, for a summary on these geometries 
and the explicit expression of the U(1)'s of interest). It should 
also be of 
interest to study different geometries, like the ones found by 
Pando-Zayas and Tseytlin \cite{PandoZayas:2000sq}. As well as other 
geometries in massive IIA; for example the
duals of CFT's in five and three dimensions that are constructed in 
\cite{Nunez:2001pt} also present all the necessary 
characteristics for the $SL(3,R)$ transformations to be 
applied. Computing observables in these transformed
geometries could be a way of understanding better the dual field 
theories.

To end this paper, we want to stress that the differences and 
similarities between the
original confining model and the transformed one leads us, by
comparison, to a better understanding of the Physics underlying
these models. 
\section{Acknowledgments}
We would like to thank Changhyun Ahn, Dan Freedman, Joanna 
Karczmarek, Ingo Kirsch,
Martin Kruczenski, 
Juan Martin Maldacena, Angel Paredes-Galan, Leo 
Pando-Zayas, Martin Schvellinger, Alfonso Ramallo for 
discussion, comments and 
correspondence that helped improving the presentation of this work. 
This work was 
supported  in part by funds
provided by the U.S.Department of Energy (DoE) under cooperative
research agreement DF-FC02-94ER408818. Carlos Nunez is a Pappalardo 
Fellow.  
\appendix
\section{Appendix: Solution Generating Technique}
\renewcommand{\theequation}{A.\arabic{equation}}
\setcounter{equation}{0}
In this appendix we summarize the technique which was used to 
generate supergravity
solutions presented in the paper \cite{Lunin:2005jy}.

Let us write the most general configuration in IIB as 
\bea
ds_{IIB}^2&=&F\left[\frac{1}{\sqrt{\Delta}}
(D\varphi_1-C(D\varphi^2))^2
+\sqrt{\Delta}(D\varphi_2)^2\right]
+ \frac{e^{2\phi/3} }{F^{1/3}}
g_{\mu\nu}dx^\mu dx^\nu,\nonumber\\
B&=&B_{12}(D\varphi^1)\wedge (D\varphi^2)+
\left\{B_{1\mu}(D\varphi^1)+B_{2\mu}(D\varphi^2)\right\}\wedge
dx^\mu
\nonumber\\
&&-\frac{1}{2}A^m_\mu B_{m\nu}dx^\mu\wedge dx^\nu+
\frac{1}{2}{\tilde b}_{\mu\nu}dx^\mu\wedge dx^\nu
\nonumber\\
e^{2\Phi}&=&e^{2\phi},\qquad C^{(0)}=\chi
\nonumber\\
C^{(2)}&=&C_{12}(D\varphi^1)\wedge (D\varphi^2)
+\left\{C_{1\mu}(D\varphi^1)+C_{2\mu}(D\varphi^2)\right\}\wedge
dx^\mu\nonumber\\
&&-\frac{1}{2}A^m_\mu C_{m\nu}dx^\mu\wedge dx^\nu+
\frac{1}{2}{\tilde c}_{\mu\nu}dx^\mu\wedge dx^\nu
\nonumber\\
C^{(4)}&=&-\frac{1}{2}(
{\tilde d}_{\mu\nu}+B_{12}{\tilde c}_{\mu\nu}-
\eps^{mn}B_{m \mu}C_{n\nu}-B_{12}A^m_{\mu}C_{m\nu})
dx^\mu dx^\nu D\varphi^1 D\varphi^2 \nonumber\\
&&+\frac{1}{6}({  C}_{\mu\nu\lam}+3({\tilde b}_{\mu\nu}+
A^1_{\mu}B_{1\nu}-A^2_{\mu}B_{2\nu})C_{1\lam}) dx^\mu dx^\nu
dx^\lam D\varphi^1 + \lab{IIBform}
\\ && + d_{\mu_1 \mu_2\mu_3\mu_4} dx^{\mu_1}  dx^{\mu_2} dx^{\mu_3} 
dx^{\mu_4} +
 \hat d_{\mu_1 \mu_2 \mu_3  } dx^{\mu_1}  dx^{\mu_2} 
dx^{\mu_3}D\varphi^2
\nonumber \lab{cfour} \eea where as defined in the main text of this 
paper \bea
D\varphi^2=d\varphi^2+A^2,\qquad D\varphi^1=d\varphi^1+{A}^1 \eea

We have three objects which transform as vectors (see also 
(\ref{transf11})-(\ref{vectors}) in the main text): \bea
&&V^{(1)}_\mu=\left(\begin{array}{c} -B_{2\mu}\\A^1_\mu\\C_{2\mu}
\end{array}\right),\quad
V^{(2)}_\mu=\left(\begin{array}{c}
B_{1\mu}\\A^2_\mu\\-C_{1\mu}
\end{array}\right):\quad V^{(i)}_\mu\rightarrow 
(\Lambda^T)^{-1}V^{(i)}_\mu;
\lab{1forms}\\
&&
W_{\mu\nu}=\left(\begin{array}{c}
{\tilde c}_{\mu\nu}\\{\tilde d}_{\mu\nu}\\
{\tilde b}_{\mu\nu}
\end{array}\right)\rightarrow \Lambda W_{\mu\nu}\lab{2forms}
\eea
and one matrix
\bea
M=g g^T ~,\quad g^T=
\left(\begin{array}{ccc}
e^{-\phi/3}F^{-1/3}&0&0\\
0&e^{-\phi/3}F^{2/3}&0\\
0&0&e^{2\phi/3}F^{-1/3}
\end{array}\right)
\left(\begin{array}{ccc}
1&B_{12}&0\\
0&1&0\\
\chi&-C_{12}+\chi B_{12}&1
\end{array}\right) \lab{demag}
\eea which transforms as \bea M\rightarrow \lambda  M\Lambda^T
\eea The scalars $\Delta, ~C$ as well as the three form ${
C}_{\mu\nu\lam}$ stay invariant under these $SL(3,R)$
transformations.
Finally, let us write the elements of the transformed matrix 
$(g^T)'$ 
for the most general transformation, with parameters $(\sigma, 
\gamma)$ nonzero and find the expression for the
transformed fields $B_{12}', C_{12}', \chi', e^{\phi'}, F' $
\bea
& & g_{11}^T= \frac{e^{-\phi/3} \kappa}{\mu}, \;\; g_{12}^{T}= 
\frac{e^{5/3\phi }}{\mu\kappa} (B_{12} +\gamma B_{12}^2 - 
B_{12}C_{12}\sigma + F^2(\gamma -\chi \sigma)),\;\;g_{2,2}^T= 
\frac{(e^{2\phi } F^2)^{1/3}}{\kappa},\nonumber\\ 
& & g_{32}^T=\frac{e^{-\phi/3}}{\mu}(B_{12}\chi e^{2\phi} + C_{12}^2 
\sigma e^{2\phi} + B_{12}^2\sigma (1+\chi^2 e^{2\phi})+ F^2\sigma 
- C_{12}e^{2\phi} (1+ 2 B_{12}\chi\sigma) )\nonumber\\
& & g_{31}^T= \frac{e^{-\phi/3}}{\mu}(- C_{12}\gamma e^{2\phi} + 
C_{12}^2\gamma \sigma e^{2\phi} + B_{12}\chi^2 e^{2\phi}\sigma(1+ 
B_{12}\gamma) +\sigma(B_{12} + \nonumber\\
& & B_{12}^2\gamma + 
F^2\gamma) -\chi 
e^{2\phi} (-1 + C_{12}\sigma + B_{12}\gamma (2 C_{12}\sigma 
-1)))\nonumber\\
& & g_{3,3}^T= (\frac{e^{-\phi}}{F})^{1/3}\sqrt{(B_{12}^2 + 
F^2)\sigma^2 + 
e^{2\phi}(1 - C_{12}\sigma + B_{12}\chi\sigma)^2}, \nonumber\\
& & \mu= 
F^{1/3}\sqrt{(B_{12}^2 + F^2)\sigma^2 + e^{2\phi} (1 - 
C_{12}\sigma + B_{12}\sigma\chi)^2} \nonumber\\
& & \kappa^2= F^2\sigma^2 + e^{2\phi}(\; (B_{12}\gamma)^2 - 2 
B_{12}\gamma(C_{12}\sigma -1) + (C_{12}\sigma-1)^2 + F^2(\gamma 
-\sigma\chi)^2)
\eea
all the other components are zero. \footnote{ We thank Changhyun Ahn 
for 
pointing out typos in a previous version of this appendix} The 
transformed fields are \bea
& & B_{12}'=\frac{g_{12}^T}{g_{11}^T}, \;\; 
e^{\phi'}=\frac{g_{33}^T}{g_{11}^T}, \;\; 
\chi'=(\frac{g_{22}^T g_{11}^T}{g_{33}^T})^{1/3}g^T_{31}\nonumber\\
& & C_{12}'= \chi' B_{12}' - g_{32}^T  g_{22}^T
 g_{11}^T
\label{a88}
\eea
Some colleagues might find useful the complete expression of the 
transformed fields in the most general situation, with $C_{12}, 
B_{12}, \chi, \phi$ turned on, that can be obtained from the 
previous eq.(\ref{a88})


\section{Appendix : The Non-Commutative ${\cal N}=1$ SYM Solution}

\renewcommand{\theequation}{B.\arabic{equation}}
\setcounter{equation}{0}

We might try the methods developed in the the core of the paper, to 
make a transformation on the directions where the gauge theory 
lives.
Let us pick these two $U(1)'s$ to be the ones labeled by the 
compactified coordinates $x_1, x_2$. the reader can check that in 
this case, the original configuration can be written as
\bea
& & ds^2_{10}\,= F (\frac{(Dx_1 - C Dx_2)^2}{\sqrt{\Delta}} + 
\sqrt{\Delta} Dx_2^2) + (\frac{e^{2/3\phi} }{F}) (e^{-2/3 
\phi} F) \alpha' g_s N 
e^{{\phi}}\,\,\Big[\,
\frac{1}{\alpha' g_s N 
}dx^2_{1,1}\,+\nonumber\\
& & e^{2h}\,\big(\,d\theta^2+\sin^2\theta 
d\varphi^2\,\big)\,+\,
dr^2\,+\,{1\over 4}\,(w^i-A^i)^2\,\Big]\,\,,
\label{metricnc}
\eea

With $F= e^{\phi}, \; \Delta=1,\;\;C=0, {\cal A}^{(i)} =0,$ and with 
our 
eight 
dimensional metric  given by
\beq
g_{\mu\nu} dx^\mu dx^\nu= (e^{-2/3
\phi} F) \alpha' g_s N
e^{{\phi}}\,\,\Big[\,
\frac{1}{\alpha' g_s N
}dx^2_{1,1}\,+\,e^{2h}\,\big(\,d\theta^2+\sin^2\theta
d\varphi^2\,\big)\,+\,
dr^2\,+\,{1\over 4}\,(w^i-A^i)^2\,\Big]\,\,,
\eeq
and the RR two form
\bea
C_{(2)}&=&{1\over 4}\,\Big[\,\psi\,(\,\sin\theta d\theta\wedge d\varphi\,-\,
\sin\tilde\theta d\tilde\theta\wedge d\tilde\varphi\,)
\,-\,\cos\theta\cos\tilde\theta d\varphi\wedge d\tilde\varphi\,-\rc\rc
&&-a\,(\,d\theta\wedge w^1\,-\,\sin\theta d\varphi\wedge 
w^2\,)\,\Big]= \tilde{c}_{\mu\nu} dx^\mu \wedge dx^\nu 
\label{RRnc}
\eea
and the rest of the fields 
\beq
{\cal A}^{(i)}= C^{(i)}= C_{12}= B_{12}= B^{(i)}= \tilde{b}= 
\tilde{d}=0
\eeq
so, upon performing the transformation in the two torus ($x_1, 
x_2$), we get a new metric
\bea
& & ds^2_{10}\,= F' (dx_1^2 + dx_2^2) + (\frac{e^{2/3\phi'} }{F'}) 
(e^{-2/3 \phi} F) \alpha' g_s N 
e^{{\phi}}\,\,\Big[\,
\frac{1}{\alpha' g_s N 
}dx^2_{1,1}\,+\nonumber\\
& &e ^{2h}\,\big(\,d\theta^2+\sin^2\theta 
d\varphi^2\,\big)\,+\,
dr^2\,+\,{1\over 4}\,(w^i-A^i)^2\,\Big]\,\,,
\label{metricncnew}
\eea
with $F'$ and the new matter fields given in (\ref{gentrans}).

So, we can see that doing the rotation in this case, has generated a 
NS two form, via the term $B_{12}'$,  and a new $C_{12}'$ (for 
nonzero $\sigma$!). If we concentrate on the $\sigma=0$ 
transformation, we see that we have to add to the metric in 
(\ref{metricncnew}) the NS two form and the RR four forms
\beq
B_2'= B_{12}' dx^1\wedge dx^2,\;\; C_2' = \tilde{c}, \;\; 2 C_4=- 
B_{12}' \tilde{c} \wedge dx^1\wedge dx^2, \;\;e^{\phi'}= 
\frac{e^{\phi}}{1+\gamma^2 e^{2\phi}}
\eeq
this is precisely the configuration found by Mateos, Pons and 
Talavera in \cite{Mateos:2002rx}.
It is quite a nice check of the method and also a check of the way 
in which we are thinking about this transformed configurations. 
Indeed, the authors in \cite{Mateos:2002rx} have checked that many 
observables do not change compared to the commutative background.

\section{Appendix: The Non-Commutative KK theory}

\renewcommand{\theequation}{C.\arabic{equation}}
\setcounter{equation}{0}

In this appendix, we will write the expression for the metric in the 
case in which we pick up the two torus to be in one of the 
directions of the field theory (that we label by $z$) and the angle 
in the two sphere, labeled before as $\varphi$, notice that this 
gives a NC six dimensional field theory for the KK modes. We could 
also choose 
the second angle to be $\tilde{\varphi}$ and in this case, like in 
previous sections the background would have been dual to a dipole 
for the KK field  theory. 

The metric reads
\bea
& & ds_{string}^2= e^{\phi} [ dx_{1,2}^2 +  \alpha' g_sN dr^2]  
+ G  d\psi^2 + D d\theta^2 + E d\tilde{\theta}^2 +  H d\theta 
d\tilde{\theta}+
+  K  d\theta d\psi  + P  d\tilde{\theta} d\psi +\nonumber\\
& &  N 
d\tilde{\varphi}^2 + Q d\tilde{\theta} d\tilde{\varphi} + 
R d\psi d\tilde{\varphi} + S d\theta d\tilde{\varphi} 
+\frac{F}{\sqrt{\Delta}} 
[d\varphi +   C dz +\alpha_1 d\theta + \alpha_2 d\tilde{\theta} 
+\alpha_3 d\psi + \alpha_4 d\tilde{\varphi} ]^2+\nonumber\\ & &
F\sqrt{\Delta}dz^2
\label{newmetricakknc}
\eea
with
\bea
& & F^2= \frac{e^{2\phi}}{\Delta}= \frac{\alpha'g_s N 
e^{2\phi}}{4}(\cos^2\theta +(4 e^{2h} + a^2)\sin^2\theta), 
\;\;\alpha_1=K=C=0, \;\;H=\frac{\alpha'g_s N
e^{2\phi}}{2}\cos\psi\nonumber\\
& & D=\frac{\alpha'g_s N e^{2\phi}}{4}(\cos^2\theta 
+(4 e^{2h} + a^2)\sin^2\theta),\;\; S= \frac{\alpha'g_s 
Ne^{2\phi}}{2}a \sin\psi \sin\tilde{\theta}, 
\nonumber\\ & & 
E= \frac{\alpha'g_s N e^{\phi}}{4}(\frac{\cos^2\theta 
+(4 e^{2h} + a^2 \cos^2\psi)\sin^2\theta}{\cos^2\theta +(4 
e^{2h} + a^2)\sin^2\theta}),\;\; G= \frac{\alpha'g_s N 
e^{\phi}}{4}(\frac{(4 e^{2h} + a^2 )\sin^2\theta}{\cos^2\theta +(4
e^{2h} + a^2)\sin^2\theta}),\;\;\nonumber\\
& & N=  \frac{\alpha'g_s N e^{\phi}}{4}\frac{4 e^{2h}\sin^2\theta + 2 a 
\cos\psi \cos\theta \sin\theta \cos\tilde{\theta} 
\sin\tilde{\theta}+ \cos^2\theta \sin^2\tilde{\theta} + a^2 
\sin^2\theta (1- \sin^2\tilde{\theta} \cos^2\psi) }{\cos^2\theta +(4
e^{2h} + a^2)\sin^2\theta},\;\;\nonumber\\
& & Q=  \frac{\alpha'g_s N e^{\phi}}{2} \sin\psi 
\sin\theta(\frac{a \cos\psi \sin\theta 
\sin\tilde{\theta} - \cos\theta \cos\tilde{\theta}}{\cos^2\theta
+(4 e^{2h} + a^2 )\sin^2\theta}),\nonumber\\ & & 
R= \frac{\alpha'g_s N 
e^{\phi}}{2} 
\sin\theta(\frac{a \cos\psi \cos\theta
\sin\tilde{\theta} + \sin\theta \cos\tilde{\theta}(4 
e^{2h} + a^2)}{\cos^2\theta
+(4 e^{2h} + a^2 )\sin^2\theta}),\nonumber\\
& &
\alpha_2=\frac{a \sin\psi \sin\theta}{\cos^2\theta +(4 
e^{2h} + 
a^2)\sin^2\theta},\;\;\alpha_3= \frac{\cos\theta}{\cos^2\theta +(4 
e^{2h} + a^2)\sin^2\theta},\nonumber\\ & &
\alpha_4=\frac{\cos\theta \cos\tilde{\theta} - a \cos\psi \sin\theta 
\sin\tilde{\theta} }{\cos^2\theta +(4 e^{2h} + a^2)\sin^2\theta}, 
\eea
the gauge and tensor  fields, following the notation adopted in the 
main text are given by
\bea
& & {\cal A}^{(1)}= \alpha_2 d\tilde{\theta} + \alpha_3 d\psi 
+\alpha_4 
d\tilde{\varphi}, \;\; {\cal A}^{(1)}=0, \nonumber\\
& & C^{(1)}=\frac{a}{4} \sin\theta \omega_2
-\frac{\psi}{4}\sin\theta d\theta 
-\frac{1}{4}\cos\theta\cos\tilde{\theta} d\tilde{\varphi},\;\; 
C^{(2)}=0\nonumber\\
& & \tilde{c}= C^{(1)}\wedge {\cal A}^{(1)} -\frac{1}{4}(\psi 
\sin\tilde{\theta} d\tilde{\theta}\wedge d\tilde{\varphi} + a 
d\theta \wedge \omega_1), \;\; C_{12}=0. 
\eea
the laws of transformation are those in 
(\ref{transf11})-(\ref{transf18}) and the reader can easily obtain 
the background corresponding to this NC KK fields theory.

\section{Appendix: Details on Rotations in R-symmetry Direction}

\renewcommand{\theequation}{D.\arabic{equation}}
\setcounter{equation}{0}

Here we give details of the computations of section 4.2. Components
of the RR two-form and the gauge connections are,
\beq\lab{D1}
C_{12}= -\frac{ \cos\tilde{\theta}}{4}, \;\; C^{(1)}= 
\frac{\cos\theta}{4} d\varphi, \;\; A^{(1)}= \cos\theta d\varphi, 
\;\;C^{(2)}= A^{(2)}=\tilde{c}_{2}=0
\eeq
The RR gauge two form can be written as
\beq
C_2= \frac{1}{4}[d\psi \wedge (\cos\theta d\varphi 
- \cos\tilde{\theta} d\tilde{\varphi})  - 
\cos\theta \cos\tilde{\theta} d\varphi \wedge d\tilde{\varphi}] 
= C_{12} D\psi \wedge D\tilde{\varphi} - 
C^{(1)}\wedge D\psi - \frac{1}{2}A^{(1)}\wedge C^{(1)}.
\eeq
After the transformation we will have a metric that reads as in 
(\ref{transformedRsym}). For completeness
let us write the 
transformed vielbein in Einstein frame 
defined by
\bea
& & e^{xi}= U  dx_i, \;\; e^r = \eta U  dr, \;\; e^\theta= \eta U  
e^{ h} d\theta, \;\; \nonumber\\
& & e^\varphi= \eta U  e^{h} \sin\theta d\varphi, \;\;e^{\tilde{\theta}}= 
\frac{\eta}{2} U  d\tilde{\theta}, \;\; e^{\tilde{\varphi}}= e^{-\phi'/4}
\sqrt{F' \sqrt{\Delta}} d\tilde{\varphi}, \nonumber\\
& & e^\psi =e^{-\phi'/4} \sqrt{\frac{F'}{\sqrt{\Delta}}} 
(d\psi + \cos\theta d\varphi + 
\cos\tilde{\theta} 
d\tilde{\varphi})
\label{vielbeint}
\eea
where $\eta =\sqrt{g_s N \alpha'}$ and we have defined, $U= e^{(\phi' + 2 
\phi)/12} (\frac{F}{F'})^{1/6}$ while the matter fields transform according to 
the rules in (\ref{gentrans}).
When NS flux is absent, the $\g$ transformation never changes the
connection one-forms ($C$ one-forms do not change for any transformation):
\beq
(C^{(1)})'=C^{(1)}=\frac{\cos\theta}{4} d\varphi, 
\;\;(A^{(1)})'=A^{(1)}=\cos\theta d\varphi.
\eeq
So, the transformed RR field is
\beq
(C_2)'= \frac{\cos\theta}{4} d\psi \wedge d\varphi + C_{12}' [d\psi + 
\cos\theta d\varphi]\wedge d\tilde{\varphi}, \;\; 
C_{12}'= \frac{C_{12}}{(F^2\gamma^2 +1)},
\eeq
and the new NS 2 form becomes
\beq
(B_{2,NS})'= B'_{12} [d\psi +
\cos\theta d\varphi]\wedge d\tilde{\varphi}, \;\;\; B_{12}'= 
\frac{F^2\gamma}{F^2\gamma^2+1}
\eeq
The new dilaton and the axion fields are,
$
e^{2\phi'}= \frac{e^{2\phi}}{F^2\gamma^2+1}, \; \chi'= \gamma C_{12}.
$
Let us also present the RR field strength,
\beq
 H_3'=  dB_2 = 
\frac{e^{-2 h+ \phi'/4}}{\eta^2 U^2 \sqrt{F'\sqrt{\Delta}}}B_{12}' 
e^\varphi \wedge e^\theta \wedge e^{\tilde{\varphi}} +
\frac{e^{\phi'/2}}{\eta F' U} (\partial_r B_{12}'  e^r + 2
\partial_{\tilde{\theta}} B_{12}' e^{\tilde{\theta}})\wedge e^\psi \wedge
e^{\tilde{\varphi}}
\label{H3p}
\eeq
\bea
& & F_3' = dC_2' -\chi' H_3'= \nonumber\\
& & \frac{e^{-2 h +\phi'/4}}{\eta^2 U^2 \sqrt{F'\sqrt{\Delta}}}( C_{12}' 
+\frac{\cos\tilde{\theta}}{4} -\chi' B_{12}') e^\varphi \wedge e^\theta 
\wedge e^{\tilde{\varphi}} + 
\frac{e^{-2 h +\phi'/4}}{4 \eta^2 U^2 \sqrt{\frac{F'}{\sqrt{\Delta}}}} 
e^\theta \wedge 
e^\varphi \wedge e^\psi \nonumber\\
& & + 
\frac{e^{\phi'/2}}{F' \eta U} ([\partial_r C_{12}' - 
\chi \partial_r B_{12}'] e^r + 2 
[\partial_{\tilde{\theta}} C_{12}' - \chi \partial_{\tilde{\theta}} 
B_{12}']e^{\tilde{\theta}})\wedge e^\psi \wedge 
e^{\tilde{\varphi}}.
\label{F3pp}
\eea

\section{Appendix: Explicit Form of the coordinate transformation in
  Section 6.4}
\renewcommand{\theequation}{F.\arabic{equation}}
\setcounter{equation}{0}

Defining the new coordinates as $\bt, \bpsi, \bvf$, one shows that 
(\ref{coordtr2}) is given by:
\bea\lab{coordtr3}
\cos\bt &=& \cos\tt\cos\q-\sin\t\sin\tt\cos(\psi+\vf),\nn\\
\cot\bpsi &=& \cos\q\cos(\psi+\vf),\\
\cos\bvf \sin\bt &=& cos\tvf\sin\tt\cos\q
+\sin\q\cos\tt\cos\tvf\cos(\psi+\vf)-\sin\q\sin\tvf\sin(\psi+\vf).
\eea

\end{document}